# *Chiral Gain–Induced Time-Reversal Symmetry Breaking in Quantum Systems*

*Mário G. Silveirinha*[(1)*], *Daigo Oue,*[(1,2,3)]

[(1)] *University of Lisbon–Instituto Superior Técnico and Instituto de Telecomunicações, Avenida Rovisco Pais, 1, 1049-001 Lisboa, Portugal*

[(2)] *RIKEN Center for Advanced Photonics, Saitama 351-0198, Japan*

[(3)] *The Blackett Laboratory, Imperial College London, London SW7 2AZ, United Kingdom*

**Abstract**

Structured light offers a powerful approach to tailor light–matter interactions in quantum systems with chiral properties. While chirality has been extensively studied in passive platforms, the role of optical gain in controlling chiral quantum dynamics remains largely unexplored. In this work, we develop a general theoretical framework to describe the dynamics of an atomic system interacting with structured gain environments, where amplification depends on the light's polarization or momentum. By quantizing the electromagnetic field in linear bianisotropic media with gain and extending the Lindblad formalism to these settings, we derive a master equation governing the atom's irreversible evolution. We show that chiral gain can break time-reversal symmetry and drive the system toward a symmetry-broken steady state with nonreciprocal properties. This effect is illustrated in detail for moving plasmonic substrates, which exhibit inherently chiral gain that selectively amplifies transitions of a given handedness. Our results establish chiral gain as a novel mechanism for engineering nonreciprocal quantum steady states.

[*] To whom correspondence should be addressed: E-mail: mario.silveirinha@tecnico.ulisboa.pt

# I. Introduction

In recent years, interactions with structured light have attracted growing attention [1]. Chiral properties are ubiquitous in nature, emerging in molecules, materials, and other low-symmetry systems [2]. These systems are highly sensitive to the polarization of light, particularly to its handedness. Notably, quantum systems such as qubits often exhibit chiral features, inherited from the intrinsic spin magnetic moment of electrons and enhanced by spin–orbit coupling [3]. This observation has sparked increasing interest in harnessing chirality as a new degree of freedom for tailoring light–matter interactions at the atomic scale, giving rise to the field of chiral quantum optics [4, 5]. For instance, it has been shown that selective coupling to specific chiral transitions of a qubit can lead to strongly asymmetric and nonreciprocal behaviors, with applications in quantum information processing [6-8]. Furthermore, recent advancements in chiral cavity quantum electrodynamics with low-dimensional materials and molecules have further highlighted the relevance chiral effects in quantum matter [9, 10].

In parallel, the study of non-Hermitian processes in photonics has gained significant momentum [11, 12, 13, 14]. In particular, the introduction of optical gain has emerged as a powerful tool for enhancing the robustness and tunability of photonic platforms, enabling not only loss compensation and amplification, but also nonreciprocal light–matter interactions [15, 16].

However, to the best of our knowledge, the possibility of using gain to control or leverage the chirality (spin degree of freedom) of quantum objects remains largely unexplored. The goal of this article is to fill this gap by developing a comprehensive theory to characterize the dynamics of quantum systems interacting with structured gain environments. Of special interest are environments in which gain depends on the polarization or momentum of light. Several examples of such systems have been recently proposed, including setups where optical pumping selectively amplifies light of a given handedness [16], and platforms based on the non-Hermitian electro-optic effect in electrically biased, low-symmetry conductors with a non-trivial Berry dipole [17, 18, 19]. Furthermore, it has been shown that moving macroscopic bodies can provide directional gain, with the amplification depending on the propagation direction of light [20, 21].

Here, we demonstrate that structured gain in these systems can selectively amplify atomic transitions that match the chirality associated with the gain mechanism. We focus our analysis on the steady-state regime, where the atomic system reaches an equilibrium with the gain environment. We show that under appropriate conditions, such structured gain can break the time-reversal symmetry of the atom and bias it toward a symmetry-broken state that may exhibit permanent magnetization or a nonreciprocal polarizability response.

To this end, we first develop a framework for quantizing the electromagnetic field in generic linear bianisotropic media with gain, extending previous formulations [22-24]. Our analysis shows that, in general, gain and dissipation must be associated with independent bosonic reservoirs, and that the decomposition of the material response into gain and dissipative contributions depends on the microscopic processes at play. We then extend the Lindblad formalism to such gain environments, deriving a quantum master equation that governs the time evolution of the atom density matrix. The coefficients of the Lindblad equation are expressed in terms of the photonic Green's function and depend on the decomposition of the material response into gain and dissipative parts.

We apply our theory to different atomic systems: (i) a qubit with non-degenerate ground and excited states, and (ii) a time-reversal symmetric qutrit with a V-shaped energy structure, where the ground state is non-degenerate and the excited states form a degenerate pair. For the V-shaped system, we demonstrate that interaction with a chiral gain environment can indeed break the time-reversal symmetry of the steady state, resulting in a population imbalance between the excited states. We illustrate the theory with two examples: a substrate with isotropic gain, and a moving plasmonic substrate. Using a quasi-static approximation, we derive closed-form analytical expressions for the excitation and decay rates in each case. In particular, we find that a moving plasmonic slab provides a chiral gain response, whereby one handedness of circular polarization experiences amplification while the opposite handedness undergoes dissipation. Additionally, for the isotropic gain substrate, we show that when the atomic transition dipole is linearly polarized, the final steady state is not unique, and retains memory of the atom's initial state.

The article is organized as follows. In Sec. II, we present a general theory for the phenomenological quantization of the electromagnetic field in linear bianisotropic media with gain. In Sec. III, we derive the reduced quantum master equation describing the dynamics of an atom with two (possibly degenerate) energy levels in a gain environment. The Lindblad coefficients are expressed in terms of the system's Green function. In Sec. IV, we apply this framework to study the steady state of an atom in a structured gain environment, showing that chiral gain can break time-reversal symmetry. Finally, Sec. V summarizes the main results.

## II. Field Quantization in Gain Systems

The quantum electrodynamics of systems with gain has been explored in several previous works, both in the context of input-output relations in systems with quantized gain media [25-27] and through macroscopic field quantization approaches [21-24, 28]. In the following, we extend the widely adopted phenomenological quantization approach to arbitrary linear bianisotropic gain systems.

### *A. Gain systems*

In our analysis, we consider a general bianisotropic linear material platform described by a 6×6 material matrix $\mathbf{M}=\mathbf{M}(\mathbf{r},\omega)$ that links the $\mathbf{D},\mathbf{B}$ macroscopic fields with the $\mathbf{E},\mathbf{H}$ fields as follows:

$$\begin{pmatrix}\mathbf{D}\\ \mathbf{B}\end{pmatrix}=\mathbf{M}\cdot\begin{pmatrix}\mathbf{E}\\ \mathbf{H}\end{pmatrix}\equiv\begin{pmatrix}\varepsilon_0\overline{\varepsilon} & \frac{1}{c}\overline{\xi}\\ \frac{1}{c}\overline{\zeta} & \mu_0\overline{\mu}\end{pmatrix}\cdot\begin{pmatrix}\mathbf{E}\\ \mathbf{H}\end{pmatrix}. \tag{1}$$

As usual, $\overline{\varepsilon}$ represents the frequency dependent permittivity tensor, $\overline{\mu}$ the permeability tensor, and $\overline{\xi},\overline{\zeta}$ the magneto-electric coupling tensors.

The non-Hermitian light matter interactions are governed by the matrix:

$$\mathbf{M}''=\frac{\mathbf{M}-\mathbf{M}^{\dagger}}{2i}, \tag{2}$$

where the dagger superscript denotes the Hermitian conjugate. The material exhibits an overall dissipative response when $\mathbf{M}''$ is a positive definite matrix in the real frequency axis. Conversely, if some or all of the eigenvalues of $\mathbf{M}''$ are negative, the material exhibits gain [29]. When the material response is purely electric, $\mathbf{M}''$ effectively reduces to the non-Hermitian part of the permittivity tensor $\mathbf{M}'' \to \frac{1}{2i}\left(\bar{\varepsilon} - \bar{\varepsilon}^{\dagger}\right)$.

The field quantization in a gain system is typically based on a decomposition of the non-Hermitian response of the form:

$$\mathbf{M}'' = \mathbf{M}''_{\mathrm{L}} + \mathbf{M}''_{\mathrm{G}}, \tag{3}$$

where $\mathbf{M}''_{\mathrm{L}}$ ($\mathbf{M}''_{\mathrm{G}}$) represents the component describing dissipative (gain) interactions [21-24]. Both $\mathbf{M}''_{\mathrm{L}}$ and $-\mathbf{M}''_{\mathrm{G}}$ are Hermitian non-negative matrices. The most straightforward decomposition of $\mathbf{M}''$ consistent with Eq. (3) is based on its spectral decomposition. Specifically, as $\mathbf{M}''$ is an Hermitian matrix, it can be written in terms of its eigenvectors ($\mathbf{v}_i$) and real-valued eigenvalues ($\Lambda_i$) as

$$\mathbf{M}'' = \sum_{\Lambda_i>0} |\Lambda_i| \mathbf{v}_i \otimes \mathbf{v}_i^* - \sum_{\Lambda_i<0} |\Lambda_i| \mathbf{v}_i \otimes \mathbf{v}_i^* . \tag{4}$$

Here, $\otimes$ represents the tensor product and it is implicit that the eigenvectors are normalized as $\mathbf{v}_i^* \cdot \mathbf{v}_j = \delta_{ij}$. The first term (second term) on the right-hand side may be identified with $\mathbf{M}''_{\mathrm{L}}$ ($\mathbf{M}''_{\mathrm{G}}$). This decomposition aligns with previous studies, which primarily focused on scalar permittivity responses ($\mathbf{M}'' \to \varepsilon''$, with $\varepsilon''$ the imaginary part of the permittivity) [21-24]. However, in our understanding, this decomposition generally lacks a strong physical basis.

For instance, consider a metamaterial composite where particles with a gain electric response ($\varepsilon'' < 0$) are mixed with particles exhibiting a dissipative response ($\varepsilon'' > 0$), where $\varepsilon''$ represents the imaginary part of the permittivity, taken as a scalar for simplicity. In principle, such a system can be modeled by an effective permittivity $\varepsilon_{\mathrm{ef}}$, describing its homogenized response. When the homogenized system is stable, meaning that all excitations eventually decay due to dissipation (i.e., dissipation mechanisms dominate gain mechanisms), the effective permittivity has a positive imaginary part, $\varepsilon''_{\rm ef} > 0$.

In this scenario, the eigenvalue decomposition of the non-Hermitian response would yield $\mathbf{M}''_{\rm G} = 0$, as the overall material response is unconditionally dissipative. This would suggest that a qubit interacting with this system would always relax to its ground state, as no gain mechanisms exist to enable population inversion. However, since the actual system consists of gain particles, this description cannot be entirely correct: the qubit's interaction with the gain components can indeed facilitate transitions to excited states. This claim is consistent with a previous work [30] that pointed out that naively applying the conventional effective-medium approach in quantum optics could lead to failure (e.g., wrongly predict the squeezing effect).

The previous discussion highlights that a decomposition of $\mathbf{M}'' = \mathbf{M}''_{\rm L} + \mathbf{M}''_{\rm G}$ based purely on the eigenvalues of the dissipative response may fail to fully capture the physics of the problem. A physically meaningful decomposition must instead rely on an understanding of the actual mechanisms responsible for gain and dissipation, rather than only on their combined influence on the overall response.

*B. Field quantization*

In Appendix A, we generalize the standard phenomenological field quantization approach [31, 32] to arbitrary bianisotropic platforms with gain described by a material matrix $\mathbf{M}$, where the non-Hermitian part of the response satisfies a decomposition of the form $\mathbf{M}'' = \mathbf{M}''_{\rm L} + \mathbf{M}''_{\rm G}$, with $\mathbf{M}''_{\rm L}$ and $-\mathbf{M}''_{\rm G}$ being Hermitian non-negative matrices. As discussed in the previous subsection, this decomposition is not necessarily determined by spectral decomposition of $\mathbf{M}''$, but rather depends on the specific physical mechanisms that govern the non-Hermitian response. We find that field quantization in this general context necessitates the introduction of two sets of independent bosonic operators: one corresponding to the dissipative part of the response ($\hat{\mathbf{b}}_{\mathrm{L},\omega}(\mathbf{r}_0)$) and the other ($\hat{\mathbf{b}}_{\mathrm{G},\omega}(\mathbf{r}_0)$) to the gain response. This distinction is necessary because, in general, gain and dissipation can coexist within the same material at the same frequency. Their anisotropic nature and competitive interactions lead to physical effects that cannot be simply described by combining both responses into a single set of bosonic operators. Furthermore, for general bianisotropic media $\hat{\mathbf{b}}_{\alpha,\omega}(\mathbf{r}_0)$ are 6 component operators (six-vector operators) ($\alpha = \mathrm{L,G}$).

The electromagnetic field Hamiltonian is expressed in terms of both positive-frequency and negative-frequency quantum harmonic oscillators as follows [Eq. (A9)]:

$$\hat{H}_{\mathrm{EM}} = \int_0^\infty d\omega \int d^3\mathbf{r}\, \hbar\omega \left[ \hat{\mathbf{b}}^\dagger_{\mathrm{L},\omega}(\mathbf{r}) \cdot \hat{\mathbf{b}}_{\mathrm{L},\omega}(\mathbf{r}) - \hat{\mathbf{b}}^\dagger_{\mathrm{G},\omega}(\mathbf{r}) \cdot \hat{\mathbf{b}}_{\mathrm{G},\omega}(\mathbf{r}) \right]. \tag{5}$$

Here, $\hat{\mathbf{b}}_{\mathrm{L},\omega}(\mathbf{r}_0)$ and $\hat{\mathbf{b}}_{\mathrm{G},\omega}(\mathbf{r}_0)$ are operators satisfying bosonic commutation relations $\left[\hat{\mathbf{b}}_{\alpha,\omega}(\mathbf{r}_0), \hat{\mathbf{b}}^\dagger_{\alpha,\omega'}(\mathbf{r}'_0)\right] = \mathbf{1}_{6\times 6}\,\delta(\omega-\omega')\,\delta(\mathbf{r}_0-\mathbf{r}'_0)$ and $\left[\hat{\mathbf{b}}_{\alpha,\omega}(\mathbf{r}_0), \hat{\mathbf{b}}_{\alpha,\omega'}(\mathbf{r}'_0)\right] = 0$. All the operators with indices L and G commute. The vacuum state $|0_{\mathrm{E}}\rangle$ is defined so that $\hat{\mathbf{b}}_{\mathrm{L},\omega}|0_{\mathrm{E}}\rangle = 0 = \hat{\mathbf{b}}_{\mathrm{G},\omega}|0_{\mathrm{E}}\rangle$.

From Eq. (5), it follows that $\hat{\mathbf{b}}_{\mathrm{L},\omega}(\mathbf{r}_0)$ ($\hat{\mathbf{b}}_{\mathrm{G},\omega}(\mathbf{r}_0)$) correspond to positive (negative) frequency quantum harmonic oscillators. Since positive (negative) frequency oscillators have spectra bounded from below (above), and both types typically coexist within the same material platform, the energy spectrum of gain systems is generally unbounded [24].

The emergence of negative frequency oscillators can be justified by noting that our Hamiltonian only describes the linearized dynamics of the quantum system. Specifically, systems with gain are operated around some driving point determined by a bias (e.g., optical pumping or electric bias). Evidently, the Hamiltonian describing the full quantum system must be positive, and typically leads to nonlinear Heisenberg equations of motion. For weak perturbations around the bias point, it is possible to linearize the system response (e.g., by treating the bias as a classical field or by applying the Holstein–Primakoff map [33]), such that the linearized Hamiltonian is described by a quadratic form involving only quantum harmonic oscillators. Different from the full Hamiltonian, the effective Hamiltonian in gain systems does not need to be positive (see an example in [34]), consistent with our result. Hence, the unbounded spectra of $\hat{H}_{\rm EM}$ is attributed to the fact that the pump mechanism associated with the optical gain is not explicitly included in the linearized quantum theory.

In particular, it follows that the "vacuum state" with no field quanta ($|0_{\rm E}\rangle$) does not minimize the field energy, as it is determined by the bias point. However, it still represents a state with no field excitations and, as such, minimizes the variance of the electromagnetic field perturbations around the bias point. This makes it the closest quantum analogue to a vanishing classical field perturbation in the presence of an external bias. Hamiltonians with an unbounded spectrum naturally arise in the electrodynamics of moving systems [28, 34].

The quantized electromagnetic fields $\hat{\mathbf{F}} = \begin{pmatrix}\hat{\mathbf{E}} & \hat{\mathbf{H}}\end{pmatrix}^T$ are written in terms of the bosonic operators as $\hat{\mathbf{F}} = \hat{\mathbf{F}}_{\mathrm{L}} + \hat{\mathbf{F}}_{\mathrm{G}} + h.c.$ with $\hat{\mathbf{F}}_{\mathrm{L}}, \hat{\mathbf{F}}_{\mathrm{G}}$ defined as [Eq. (A11)] ($h.c.$ stands for the Hermitian conjugate operator):

$$\hat{\mathbf{F}}_{\mathrm{L}}\left(\mathbf{r},t\right) = \int_0^{\infty} d\omega\, e^{-i\omega t} \sqrt{\frac{\hbar}{\pi|\omega|}} \int d^3\mathbf{r}_0\, \overline{\mathcal{G}}\left(\mathbf{r},\mathbf{r}_0,\omega\right)\cdot \mathbf{R}_{\mathrm{L}\omega}\left(\mathbf{r}_0\right)\cdot \hat{\mathbf{b}}_{\mathrm{L},\omega}\left(\mathbf{r}_0\right), \tag{6a}$$

$$\hat{\mathbf{F}}_{\mathrm{G}}\left(\mathbf{r},t\right) = \int_0^{\infty} d\omega\, e^{-i\omega t} \sqrt{\frac{\hbar}{\pi|\omega|}} \int d^3\mathbf{r}_0\, \overline{\mathcal{G}}\left(\mathbf{r},\mathbf{r}_0,\omega\right)\cdot \mathbf{R}_{\mathrm{G}\omega}\left(\mathbf{r}_0\right)\cdot \hat{\mathbf{b}}^{\dagger}_{\mathrm{G},\omega}\left(\mathbf{r}_0\right). \tag{6b}$$

Here, $\overline{\mathcal{G}}\left(\mathbf{r},\mathbf{r}_0,\omega\right)$ is the system's Green's function defined in Appendix A [Eq. (A2)], and $\mathbf{R}_{\mathrm{L}\omega} = \left[\omega \mathbf{M}''_{\mathrm{L}}\right]^{1/2}$ and $\mathbf{R}_{\mathrm{G}\omega} = \left[-\omega \mathbf{M}''_{\mathrm{G}}\right]^{1/2}$ represent the contributions from dissipative and gain processes. In Appendix A, we demonstrate that the quantized electromagnetic fields satisfy the expected canonical commutation relations [Eqs. (A19) and (A21)], ensuring the consistency of the quantization framework.

It is important to underline that in our construction it is essential to model the gain channels as negative-energy harmonic oscillators, as in Eq. (5). This is not a choice of convenience: it is essential to ensure the correct algebraic structure of the quantum theory. A Hamiltonian composed only of positive-energy oscillators would not reproduce the standard field commutation relations in the presence of gain.

### *C. Generalized Gibbs state and fluctuation-dissipation relations*

The equilibrium state of a standard passive systems is the so-called Gibbs state, which is determined by the environment's density matrix $\hat{\rho}_{\mathrm{E}} = \dfrac{e^{-\beta \hat{H}}}{Z}$, where $\beta = 1/k_B T$ the inverse temperature, $k_B$ is the Boltzmann constant, $T$ is the system temperature and $Z = \mathrm{tr}\left\{e^{-\beta \hat{H}}\right\}$ is the partition function.

As discussed in the previous subsection, the Hamiltonian in gain systems is unbounded. If one were to apply the conventional definition of the Gibbs state in this context, it would preferentially populate the states with large quantum numbers of the negative-energy harmonic oscillators.

Despite considering systems with gain, we assume their overall response remains stable. Specifically, we require that the system's Green's function $\overline{\mathcal{G}}(\mathbf{r},\mathbf{r}_0,\omega)$ be free of singularities in the upper-half frequency plane, ensuring that exponentially growing oscillations, such as those in a laser, are not permitted. In fact, the gain platforms studied in this article exhibit an overall dissipative response, with $\mathbf{M}''$ strictly positive definite in the real-frequency axis. In classical terms, this means that when external excitations are switched off, the free-field oscillations of the electromagnetic field are damped, ultimately leading to a vanishing field at large times. Consequently, the equilibrium state of the corresponding quantum system should favor the lowest quantum number states.

However, as already noted, the conventional Gibbs state definition conflicts with this expectation, as it favors the occupation of negative-energy oscillator states with large quantum numbers. To resolve this inconsistency, we introduce a generalized Gibbs state, defined as:

$$\hat{\rho}_{\mathrm{E,Gibbs}} = \frac{e^{-\left(\beta_{\mathrm{L}}\hat{H}_{\mathrm{EM,L}} + \beta_{\mathrm{G}}\hat{H}_{\mathrm{EM,G}}\right)}}{Z}, \qquad Z = \mathrm{tr}\left\{e^{-\left(\beta_{\mathrm{L}}\hat{H}_{\mathrm{EM,L}} + \beta_{\mathrm{G}}\hat{H}_{\mathrm{EM,G}}\right)}\right\}, \tag{7}$$

with

$$\hat{H}_{\text{EM},\alpha} = s_\alpha \int_0^\infty d\omega \int d^3\mathbf{r}\, \hbar\omega\, \hat{\mathbf{b}}^\dagger_{\alpha,\omega}(\mathbf{r}) \cdot \hat{\mathbf{b}}_{\alpha,\omega}(\mathbf{r}), \qquad (s_{\text{L}} = 1 \text{ and } s_{\text{G}} = -1) \tag{8}$$

the positive and negative energy parts of the Hamiltonian, defined such that $\hat{H}_{\text{EM}} = \hat{H}_{\text{EM,L}} + \hat{H}_{\text{EM,G}}$ ($\alpha = \text{L,G}$). In the above, $\beta_{\text{L}}$ ($\beta_{\text{G}}$) represents the inverse (effective) temperature of the positive (negative) energy oscillators. Thus, the inverse temperature is allowed to vary across distinct subsets of harmonic modes. Related concepts have been discussed previously in Refs. [22, 35-38].

Importantly, for the generalized Gibbs state to be physically meaningful (favoring states with low occupation numbers) it is necessary that $\beta_{\text{L}} > 0$ and $\beta_{\text{G}} < 0$. This implies that the negative-energy oscillators must effectively experience a negative temperature, typical of population-inverted systems.

In principle, $\beta_{\text{L}}$ and $\beta_{\text{G}}$ should be regarded as independent parameters. In fact, $\beta_{\text{G}}$ is expected to correlate with the strength of the gain response and to depend on the specific gain mechanism. Importantly, we demonstrate in Appendix B that in the particular case where $\beta_{\text{L}} = -\beta_{\text{G}} \equiv \beta$, the field correlations are governed by standard Bose Einstein statistics described by a single (positive) inverse effective temperature $\beta = 1/k_B T$. Indeed, the generalized Gibbs state with $\beta_{\text{L}} = -\beta_{\text{G}} \equiv \beta$ gives rise to generalized fluctuation-dissipation relations whose strength is governed by the coefficient $N_\omega + \frac{1}{2}$, where $N_\omega = \frac{1}{e^{+\beta\hbar\omega} - 1}$ represents the occupation number for a quantum harmonic oscillator with frequency $\omega$. Specifically, the field correlations for the generalized Gibbs state with $\beta_{\text{L}} = -\beta_{\text{G}} \equiv \beta$ are determined by the following unilateral spectral density [Eq. (B12)]:

$$\left\langle \left\{ \hat{\mathbf{F}}(\mathbf{r})\hat{\mathbf{F}}(\mathbf{r}') \right\}_+ \right\rangle_{T,\omega}$$

$$= \frac{2\hbar}{\pi}\left( N_\omega + \frac{1}{2} \right) \mathrm{Re} \int d^3\mathbf{r}_0 \, \overline{\mathcal{G}}(\mathbf{r},\mathbf{r}_0,\omega) \cdot \left[ \mathbf{M}''_{\mathrm{L}}(\mathbf{r}_0) - \mathbf{M}''_{\mathrm{G}}(\mathbf{r}_0) \right] \cdot \overline{\mathcal{G}}(\mathbf{r}',\mathbf{r}_0,\omega)^\dagger \tag{9}$$

The element $(i,j)$ of the left-hand side tensor represents the symmetrized operator product $\hat{F}_i(\mathbf{r})\hat{F}_j(\mathbf{r}') + \hat{F}_j(\mathbf{r}')\hat{F}_i(\mathbf{r})$. For a passive system, $\mathbf{M}''_{\mathrm{G}} = 0$, we recover the standard fluctuation dissipation relations [41-42]:

$$\left\langle \left\{ \hat{\mathbf{F}}(\mathbf{r})\hat{\mathbf{F}}(\mathbf{r}') \right\}_+ \right\rangle_{T,\omega} = \left( N_\omega + \frac{1}{2} \right) \mathrm{Re} \left\{ \frac{\hbar}{i\pi} \left( \overline{\mathcal{G}}(\mathbf{r},\mathbf{r}',\omega) - \left[ \overline{\mathcal{G}}(\mathbf{r}',\mathbf{r},\omega) \right]^\dagger \right) \right\}. \tag{10}$$

This result is a consequence of Eq. (A7) in Appendix A.

Thus, a generalized Gibbs state with $\beta_{\mathrm{L}} = -\beta_{\mathrm{G}} \equiv \beta$ represents a minimal and physically reasonable choice for modeling the environment using a single effective temperature. This form aligns with earlier studies [22], which showed that the gain spectral regions effectively experience a temperature that differs in sign from the temperature of the dissipative spectral regions. Our formulation extends this idea by considering a more general scenario in which gain and dissipation coexist within the same spectral region. Moreover, this macroscopic treatment is consistent with commonly used theoretical approaches to non-equilibrium photonic systems, which often rely on a single effective temperature to characterize fluctuations and dissipation [39, 40].

The physical significance of the generalized Gibbs state with $\beta_{\mathrm{L}} = -\beta_{\mathrm{G}} \equiv \beta$ is further reinforced by the results of Refs. [44-45], which show that the generalized fluctuation-dissipation relation in Eq. (9) predicts a frictional force in moving systems that is consistent with calculations obtained through alternative theoretical approaches [21, 43], which do not rely on field quantization in gain systems. As already mentioned in the introduction (Sect. I), moving platforms represent a particular instance of a gain system.

For simplicity, throughout the remainder of this work, we model the environment using a single effective temperature such that $\beta_{\mathrm{L}} = -\beta_{\mathrm{G}} \equiv \beta$. It is relevant to note that in the zero-temperature limit, $\beta \to +\infty$, the generalized Gibbs state becomes the "vacuum state" $\hat{\rho}_{\mathrm{E,Gibbs}} \to |0_{\mathrm{E}}\rangle\langle 0_{\mathrm{E}}|$. As in the case of a passive system, increasing the temperature $T$ (i.e., decreasing $\beta$) leads to a broader occupation of photonic states and an increase of the steady-state energy stored in the environment. This aligns with the physical intuition that a quantized system should behave analogously to its classical counterpart, tending to relax into an equilibrium state with less excitations as the temperature decreases.

It is important to emphasize that in systems with gain, a true thermodynamic equilibrium is not established due to continuous external energy input. Accordingly, the temperature $T$ should be interpreted as an *effective* or phenomenological parameter characterizing the steady-state properties of the environment, rather than a fundamental thermodynamic quantity.

## III. Quantum Master Equation

In this article, we focus on the interaction of an elementary atomic system with a gain platform. The atom's interaction with the environment can be effectively modeled using a reduced quantum master equation, which typically takes the Lindblad form [46-52]. The Lindblad operator is usually derived under the Born and Markov approximations, though alternative approaches incorporating the effects of ultra-strong coupling have also been discussed in the literature [53-54]. While the Lindblad formalism has been previously employed phenomenologically in the study of gain systems [55-59], to the best of our knowledge, a rigorous first-principles derivation of the Lindblad operator—that explicitly connects the relevant coefficients with the system's geometry and material response—has not been reported.

In Appendices C and D, we provide a systematic first-principles derivation of the Lindblad operator for arbitrary linear bianisotropic platforms under the Born, Markov and secular approximations. Specifically, we assume that the environment is weakly coupled to the atom, ensuring that the environment's density matrix remains of the form $\hat{\rho}_{\mathrm{E}} \approx \hat{\rho}_{\mathrm{E,Gibbs}}$ at all times. As in Sect. II, we assume the environment is "stable", meaning that instabilities due to gain effects are excluded.

*A. Atom model*

We consider an atomic system with only two (potentially degenerate) energy levels. The ground states are denoted by $|g_i\rangle$ and the excited states by $|e_j\rangle$, with *i*=1,…, and *j*=1,…, running over the possibly degenerate ground and excited states. The ground and excited states are separated by the energy gap $\hbar\omega_{\mathrm{a}}$ so that the atom Hamiltonian is $\hat{H}_{\mathrm{at}} = \hbar\omega_{\mathrm{a}} \sum_j |e_j\rangle\langle e_j|$. Here, $\omega_{\mathrm{a}}$ is the transition frequency between the two energy levels.

Furthermore, in the Schrödinger picture, the atom transition dipole moment $\hat{\mathbf{p}}_{\mathrm{e}}$ is assumed to take the form

$$\hat{\mathbf{p}}_{\mathrm{e}}^{-} = \sum_{i,j} \boldsymbol{\gamma}_{\mathrm{e},ij} |g_i\rangle\langle e_j|, \qquad \hat{\mathbf{p}}_{\mathrm{e}}^{+} = \left(\hat{\mathbf{p}}_{\mathrm{e}}^{-}\right)^{\dagger}, \tag{11}$$

where $\boldsymbol{\gamma}_{\mathrm{e},ij}$ represents the transition electric dipole moment associated with the electronic transition $|e_j\rangle \rightarrow |g_i\rangle$. For convenience, we introduce the six-component operators $\hat{\mathbf{p}}^{\pm} = \left(\hat{\mathbf{p}}_{\mathrm{e}}^{\pm} \quad \mathbf{0}\right)^T$, which will be utilized in the formulation of the master equation below.

*B. Lindblad dissipator*

From the analysis in Appendices C and D, the time evolution of the atomic state is governed by the following reduced quantum master equation:

$$\partial_t \hat{\rho}_{\mathrm{S}} = -i\frac{1}{\hbar}\left[\hat{H}_{\mathrm{at}} + \hat{H}_{\mathrm{int,ef}}, \hat{\rho}_{\mathrm{S}}\right] + \mathcal{L}_{\mathrm{d}}\hat{\rho}_{\mathrm{S}}. \tag{12}$$

Here, $\hat{\rho}_{\mathrm{S}}$ is the density matrix of the atom, $\hat{H}_{\mathrm{at}}$ is the atom Hamiltonian, and $\hat{H}_{\mathrm{int,ef}}$ and $\mathcal{L}_{\mathrm{d}}$ are operators that describe the conservative and non-conservative interactions with the environment, respectively. For simplicity, throughout the remainder of this article, we neglect the effect of $\hat{H}_{\mathrm{int,ef}}$, which is typically associated with spectral (Lamb) shifts (the explicit formula for $\hat{H}_{\mathrm{int,ef}}$ in the zero-temperature limit is given by Eq. (C15)).

On the other hand, the dissipator $\mathcal{L}_{\mathrm{d}}$ has the Lindblad form and describes both the effect of the dissipative and gain interactions. When the environment is characterized by a generalized Gibbs state, it can be written explicitly as [Eq. (D1)]:

$$\begin{aligned}\mathcal{L}_{\mathrm{d}}\hat{\rho}_{\mathrm{S}} = &-\frac{1}{\hbar}\left\{\hat{\mathbf{p}}^{+}\cdot\overline{\boldsymbol{\mathcal{G}}}_{\mathrm{L,th}}^{\mathrm{NH}}\cdot\hat{\mathbf{p}}^{-} + \hat{\mathbf{p}}^{-}\cdot\left[\overline{\boldsymbol{\mathcal{G}}}_{\mathrm{G,th}}^{\mathrm{NH}}\right]^T\cdot\hat{\mathbf{p}}^{+}, \hat{\rho}_{\mathrm{S}}(t)\right\}_{+} \\ &+\frac{2}{\hbar}\hat{\mathbf{p}}^{-}\cdot\left[\overline{\boldsymbol{\mathcal{G}}}_{\mathrm{L,th}}^{\mathrm{NH}}\right]^T\hat{\rho}_{\mathrm{S}}(t)\cdot\hat{\mathbf{p}}^{+} + \frac{2}{\hbar}\hat{\mathbf{p}}^{+}\cdot\overline{\boldsymbol{\mathcal{G}}}_{\mathrm{G,th}}^{\mathrm{NH}}\hat{\rho}_{\mathrm{S}}(t)\cdot\hat{\mathbf{p}}^{-}.\end{aligned} \tag{13a}$$

$$\overline{\boldsymbol{\mathcal{G}}}_{\mathrm{L,th}}^{\mathrm{NH}} = \left(1 + N_{\omega_{\mathrm{a}}}\right)\overline{\boldsymbol{\mathcal{G}}}_{\mathrm{L}}^{\mathrm{NH}} + N_{\omega_{\mathrm{a}}}\overline{\boldsymbol{\mathcal{G}}}_{\mathrm{G}}^{\mathrm{NH}}, \tag{13b}$$

$$\overline{\boldsymbol{\mathcal{G}}}_{\mathrm{G,th}}^{\mathrm{NH}} = \left(1 + N_{\omega_{\mathrm{a}}}\right)\overline{\boldsymbol{\mathcal{G}}}_{\mathrm{G}}^{\mathrm{NH}} + N_{\omega_{\mathrm{a}}}\overline{\boldsymbol{\mathcal{G}}}_{\mathrm{L}}^{\mathrm{NH}}. \tag{13c}$$

Here, $\{\,,\}_+$ stands for the anti-commutator of two operators ($\{\hat{A},\hat{B}\}_+ = \hat{A}\hat{B}+\hat{B}\hat{A}$). The coefficients of the Lindblad dissipator are determined by $N_\omega = \frac{1}{e^{+\beta\hbar\omega}-1}$ evaluated at the atomic transition frequency $\omega_a$ and by the Hermitian non-negative 6×6 tensors $\overline{\boldsymbol{\mathcal{G}}}_L^{NH}$ and $\overline{\boldsymbol{\mathcal{G}}}_G^{NH}$ that govern the dissipative and gain responses of the environment. They are defined as [Eq. (C20)]:

$$\overline{\boldsymbol{\mathcal{G}}}_L^{NH} = +\int d^3\mathbf{r}'\,\overline{\boldsymbol{\mathcal{G}}}(\mathbf{r}_a,\mathbf{r}',\omega_a)\cdot\mathbf{M}''_L(\mathbf{r}')\cdot\overline{\boldsymbol{\mathcal{G}}}^\dagger(\mathbf{r}_a,\mathbf{r}',\omega_a), \tag{14a}$$

$$\overline{\boldsymbol{\mathcal{G}}}_G^{NH} = -\int d^3\mathbf{r}'\,\overline{\boldsymbol{\mathcal{G}}}(\mathbf{r}_a,\mathbf{r}',\omega_a)\cdot\mathbf{M}''_G(\mathbf{r}')\cdot\overline{\boldsymbol{\mathcal{G}}}^\dagger(\mathbf{r}_a,\mathbf{r}',\omega_a), \tag{14b}$$

with $\mathbf{r}_a$ the atom's coordinates.

From the structure of the dissipator in Eq. (13), it is evident that the terms controlled by the tensor $(1+N_{\omega_a})\overline{\boldsymbol{\mathcal{G}}}_L^{NH} + N_{\omega_a}\overline{\boldsymbol{\mathcal{G}}}_G^{NH}$ favor transitions to the atom ground states, while the terms associated with the tensor $(1+N_{\omega_a})\overline{\boldsymbol{\mathcal{G}}}_G^{NH} + N_{\omega_a}\overline{\boldsymbol{\mathcal{G}}}_L^{NH}$ promote transitions to the excited states. As expected, in the zero-temperature limit, corresponding to $N_{\omega_a} \to 0$, transitions to the excited states are only possible if the system exhibits gain ($\overline{\boldsymbol{\mathcal{G}}}_G^{NH} \neq 0$). However, at finite temperature, such transitions become feasible even in passive systems ($\overline{\boldsymbol{\mathcal{G}}}_G^{NH} = 0$) due to thermal excitations, which provide the necessary energy for population transfer to the excited states.

From Eq. (14) one sees that the tensors $\overline{\boldsymbol{\mathcal{G}}}_L^{NH}$ and $\overline{\boldsymbol{\mathcal{G}}}_G^{NH}$ are fully determined by the dissipative/gain response of the system and by the system's Green's function evaluated at the atomic transition frequency. In section IV, we will present explicit expressions for these tensors in plasmonic-type systems under a quasi-static approximation.

To conclude, we note that although this article focuses on a single two-level system, the Lindblad operator can be readily extended to qubit networks, where each atom may be a multi-level system (see Appendix C, Sect. A).

## IV. Atom Steady-State in an Environment with Gain

In the following, we apply the theory developed in the preceding sections to study the interaction of an atom with a gain environment, with particular focus on the characterization of the steady-state regime ($t \to \infty$) and the emergence of broken time-reversal symmetry due to gain mechanisms. Time-reversal symmetry breaking can influence the atom steady-state properties, such as the expectation value of its spin magnetic moment, as well as its response to external fields, for example, through changes in its atomic polarizability.

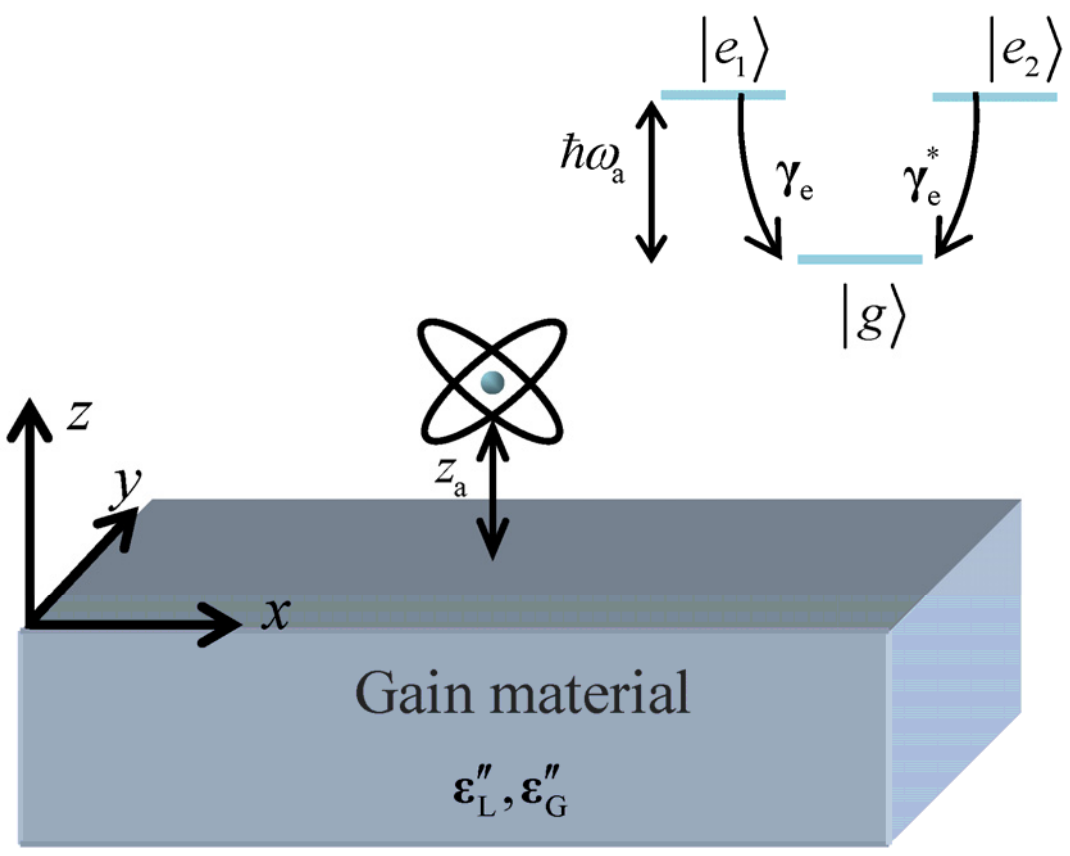

**Fig. 1** An atomic system is positioned at a distance $z_a$ from a gain dielectric substrate. The dissipative (gain) response is governed by $\varepsilon''_L$ and $\varepsilon''_G$. The inset in the top-right corner depicts the energy levels for a system with a V-shaped energy structure.

We consider the interaction of an atom with two types of environments (see Fig. 1 for an illustrative geometry): a generic isotropic substrate whose dielectric response includes a gain component, and a moving plasmonic surface. The latter is of particular theoretical interest, as it exhibits non-Hermitian interactions, either gain or dissipation, that are locked to the direction of wave propagation [21]. We will show that this gain-momentum locking leads to a chiral-gain response, where gain interactions are activated for one specific handedness of circular polarization, while dissipative interactions are triggered for the opposite handedness. Chiral-gain responses have been previously discussed in other contexts, particularly in systems involving low-symmetry metals biased by an external electric field [17-19].

*A. Non-degenerate two-level system*

To begin with, we consider the simplest case in which the atom is modeled as a two-level system with a single (non-degenerate) ground state and a single excited state (a qubit). In this scenario, the electric dipole moment operator ($\hat{\mathbf{p}}_{\rm e} = \hat{\mathbf{p}}_{\rm e}^{-} + h.c.$) is determined by $\hat{\mathbf{p}}_{\rm e}^{-} = \boldsymbol{\gamma}_{\rm e}\boldsymbol{\sigma}_{-}$ with $\boldsymbol{\sigma}_{-} = |g\rangle\langle e|$. It is convenient to introduce the rate constants:

$$\Gamma_\alpha = \frac{2}{\hbar}\boldsymbol{\gamma}_{\rm e}^{*} \cdot \overline{\boldsymbol{\mathcal{G}}}_{\alpha,{\rm th}}^{\rm NH} \cdot \boldsymbol{\gamma}_{\rm e}, \qquad \text{with } \alpha = {\rm L,G}. \tag{15}$$

Here, $\overline{\boldsymbol{\mathcal{G}}}_{\alpha,{\rm th}}^{\rm NH}$ should be identified with the electric-part of the Green's function (3×3 tensor). The Lindblad super-operator, expressed in a basis with coordinates $\hat{\rho}_{\rm S} \to \boldsymbol{\rho} = \begin{pmatrix}\rho_{\rm gg} & \rho_{\rm ge} & \rho_{\rm eg} & \rho_{\rm ee}\end{pmatrix}^{T}$, is written in terms of $\Gamma_\alpha$ as follows:

$$\mathcal{L} \rightarrow \boldsymbol{\mathcal{L}} = \begin{pmatrix} -\Gamma_{\mathrm{G}} & 0 & 0 & +\Gamma_{\mathrm{L}} \\ 0 & -\frac{\Gamma_{\mathrm{L}}}{2} - \frac{\Gamma_{\mathrm{G}}}{2} + i\omega_{\mathrm{a}} & 0 & 0 \\ 0 & 0 & -\frac{\Gamma_{\mathrm{L}}}{2} - \frac{\Gamma_{\mathrm{G}}}{2} - i\omega_{\mathrm{a}} & 0 \\ \Gamma_{\mathrm{G}} & 0 & 0 & -\Gamma_{\mathrm{L}} \end{pmatrix}. \tag{16}$$

This matrix includes both conservative and non-conservative terms, so that the time evolution of the density matrix is described by $\partial_t \boldsymbol{\rho} = \boldsymbol{\mathcal{L}} \cdot \boldsymbol{\rho}$. The steady-state regime ($t \rightarrow \infty$) is characterized by the null space of the Lindblad super-operator. It is of the form $\boldsymbol{\rho} \sim \begin{pmatrix} \Gamma_{\mathrm{L}} & 0 & 0 & \Gamma_{\mathrm{G}} \end{pmatrix}^T$. Thus, the gain environment shapes the qubit dynamics such that its long-time state is the mixed state:

$$\hat{\rho}_{\mathrm{S},\infty} = \frac{\Gamma_{\mathrm{L}}}{\Gamma_{\mathrm{L}} + \Gamma_{\mathrm{G}}} |g\rangle\langle g| + \frac{\Gamma_{\mathrm{G}}}{\Gamma_{\mathrm{L}} + \Gamma_{\mathrm{G}}} |e\rangle\langle e|. \tag{17}$$

This result is rather general and holds independently of the specific mechanism responsible for the gain or the detailed structure of the tensors $\overline{\boldsymbol{\mathcal{G}}}_{\alpha}^{\mathrm{NH}}$.

To illustrate the discussion, we consider a specific case in which the qubit interacts with a semi-infinite isotropic material exhibiting a gain component in its dielectric response (region $z < 0$). The qubit is positioned at a distance $z_{\mathrm{a}}$ above the substrate, in the air region (Fig. 1). The non-Hermitian response of the substrate is governed by the imaginary part $\varepsilon''$ of its dielectric permittivity. Following the discussion in Sect. II, this non-Hermitian response can be decomposed into dissipative and gain components: $\varepsilon''(\omega) = \varepsilon''_{\mathrm{L}}(\omega) + \varepsilon''_{\mathrm{G}}(\omega)$, with $\varepsilon''_{\mathrm{L}} > 0$ and $\varepsilon''_{\mathrm{G}} < 0$. The specific dispersion characteristics of these components depend on the underlying physical processes responsible for dissipation and amplification. Typically, material stability requires that $|\varepsilon''_{\mathrm{G}}| < \varepsilon''_{\mathrm{L}}$.

In Appendix E, we calculate the tensors $\overline{\boldsymbol{\mathcal{G}}}_{\alpha}^{\mathrm{NH}}$ using a quasi-static approximation, which neglects retardation effects by effectively assuming an infinite speed of light [Eq. (E10)]:

$$\overline{\boldsymbol{\mathcal{G}}}_{\mathrm{L,G}}^{\mathrm{NH}} = \frac{\left|\varepsilon''_{\mathrm{L,G}}\right|}{\left|\varepsilon+1\right|^2} \frac{1}{16\pi\varepsilon_0 z_{\mathrm{a}}^3} \left(\mathbf{1}_t + 2\hat{\mathbf{z}}\otimes\hat{\mathbf{z}}\right), \tag{18}$$

with $\mathbf{1}_t = \hat{\mathbf{x}}\otimes\hat{\mathbf{x}} + \hat{\mathbf{y}}\otimes\hat{\mathbf{y}}$ and the permittivity function evaluated at the atomic transition frequency ($\omega_{\mathrm{a}}$). Note that $\overline{\boldsymbol{\mathcal{G}}}_{\mathrm{L,G}}^{\mathrm{NH}}$ is positive definite as it should be. This approximation is valid for near-field interactions and has been shown to accurately capture the full electromagnetic response in different scenarios, particularly for metallic systems supporting surface plasmon polaritons (SPPs) [21, 60, 61]. The interaction tensors $\overline{\boldsymbol{\mathcal{G}}}_{\mathrm{L,G}}^{\mathrm{NH}}$ are greatly enhanced when $\mathrm{Re}\{\varepsilon\} = -1$, corresponding to the SPP resonance.

Since $\overline{\boldsymbol{\mathcal{G}}}_{\mathrm{L,G}}^{\mathrm{NH}}$ is proportional to $\left|\varepsilon''_{\mathrm{L,G}}\right|$, it follows that in the zero-temperature limit ($N_{\omega_{\mathrm{a}}} = 0$) the population of the excited state is $\rho_{\mathrm{ee}} = \dfrac{\left|\varepsilon''_{\mathrm{G}}\right|}{\varepsilon''_{\mathrm{L}} + \left|\varepsilon''_{\mathrm{G}}\right|}$. For large temperatures ($N_{\omega_{\mathrm{a}}} >> 1$) the population of the excited and ground states become equal, $\rho_{ii} \to \dfrac{1}{2}$, independent of the material's gain strength.

Typical atomic systems are time-reversal invariant. As such, non-degenerate atomic states must themselves be invariant under time reversal. Consequently, the long-time state of the two-level qubit also respects time-reversal symmetry [Eq. (17)]. This implies that signatures of nonreciprocal response, such as permanent magnetism in the steady state or nonreciprocal atomic polarizability, cannot be observed in the system under consideration, regardless of the gain profile. In the next subsection, we consider a more complex atomic structure that enables the exploration of how gain may facilitate the breaking of time-reversal symmetry in the steady-state regime.

*B. V-shaped energy structure*

Next, we consider an atomic system with a V-shaped energy structure, consisting of a nondegenerate ground state and two degenerate excited states, as illustrated in the inset of Fig. 1 (a qutrit). We assume that the excited states, which form an orthogonal basis of the excited manifold, are connected by the time-reversal symmetry operator $\mathcal{T}$, so that $|e_2\rangle = \mathcal{T}|e_1\rangle$ and $|e_1\rangle = \mathcal{T}|e_2\rangle$. For instance, a similar configuration arises in hydrogen, where the ground state $|1S\rangle$ includes a spin singlet configuration ($F$=0) that is strictly nondegenerate (note that the total spin of hydrogen is integer and thereby $\mathcal{T}^2 = +\mathbf{1}$). In the first excited manifold $|2P\rangle$ contains two excited sublevels, $|F=1, m_F=-1\rangle$ and $|F=1, m_F=+1\rangle$, which are degenerate and related by time-reversal symmetry. These two excited states carry equal and opposite total magnetic moments.

For a V-shaped energy structure, the electric dipole moment operator $\hat{\mathbf{p}}_\mathrm{e} = \hat{\mathbf{p}}_\mathrm{e}^- + h.c.$ is defined by

$$\hat{\mathbf{p}}_\mathrm{e}^- = \boldsymbol{\gamma}_\mathrm{e}|g\rangle\langle e_1| + \boldsymbol{\gamma}_\mathrm{e}^*|g\rangle\langle e_2| \equiv \boldsymbol{\gamma}_\mathrm{e}\boldsymbol{\sigma}_{1-} + \boldsymbol{\gamma}_\mathrm{e}^*\boldsymbol{\sigma}_{2-}. \tag{19}$$

Due to time-reversal symmetry, the dipole moments for the two transitions are related to each other by complex conjugation.

The Lindblad dissipator [Eq. (13)] can now be expressed as:

$$\begin{aligned}\mathcal{L}_\mathrm{d}\hat{\rho}_\mathrm{S} = &-\left\{\frac{1}{2}\sum_{ij}\Gamma_{\mathrm{L},ij}\boldsymbol{\sigma}_{i+}\boldsymbol{\sigma}_{j-} + \frac{1}{2}\sum_{ij}\Gamma_{\mathrm{G},ij}\boldsymbol{\sigma}_{j-}\boldsymbol{\sigma}_{i+}, \hat{\rho}_\mathrm{S}(t)\right\}_+ \\ &+\sum_{ij}\Gamma_{\mathrm{L},ij}\boldsymbol{\sigma}_{j-}\hat{\rho}_\mathrm{S}\boldsymbol{\sigma}_{i+} + \sum_{ij}\Gamma_{\mathrm{G},ij}\boldsymbol{\sigma}_{i+}\hat{\rho}_\mathrm{S}\boldsymbol{\sigma}_{j-}\end{aligned} \tag{20a}$$

$$\Gamma_{\alpha,ij} = \frac{2}{\hbar} \boldsymbol{\gamma}_i^* \cdot \overline{\boldsymbol{\mathcal{G}}}_{\alpha,\text{th}}^{\text{NH}} \cdot \boldsymbol{\gamma}_j, \qquad \boldsymbol{\gamma}_1 = \boldsymbol{\gamma}_\text{e}, \ \boldsymbol{\gamma}_2 = \boldsymbol{\gamma}_\text{e}^*, \tag{20b}$$

with $\boldsymbol{\sigma}_{i+} = (\boldsymbol{\sigma}_{i-})^\dagger$, $\alpha = \text{L,G}$ and $i, j = 1,2$. As the tensors $\overline{\boldsymbol{\mathcal{G}}}_{\alpha,\text{th}}^{\text{NH}}$ are Hermitian, the rate constants satisfy the symmetry relations $\Gamma_{\alpha,ij} = \Gamma_{\alpha,ji}^*$. In particular, $\Gamma_{\alpha,ii}$ is real-valued. The rates $\Gamma_{\text{L},ij}$ describe interactions that lead to downward atomic transitions, whereas the rates $\Gamma_{\text{G},ij}$ describe interactions that lead to upward transitions.

Similar to the previous subsection, the Lindblad superoperator (9×9 matrix) can be expressed in terms of the rate constants, although its full form is omitted here for brevity. As before, the steady-state is determined by the null space of the Lindblad super-operator. In the general case, the null space is non-degenerate. The corresponding atom density matrix has the following non-vanishing elements:

$$\begin{aligned}
\rho_\text{gg} &= \frac{A}{A+B}, \\
\rho_\text{e1e1} &= \frac{1}{A+B} \frac{\left(\Gamma_{\text{G},11} - \Gamma_{\text{G},22}\right) A + \Gamma_{\text{L},22} B}{\Gamma_{\text{L},11} + \Gamma_{\text{L},22}}, \\
\rho_\text{e2e2} &= \frac{1}{A+B} \frac{\left(\Gamma_{\text{G},22} - \Gamma_{\text{G},11}\right) A + \Gamma_{\text{L},11} B}{\Gamma_{\text{L},11} + \Gamma_{\text{L},22}}, \\
\rho_\text{e1e2} &= \rho_\text{e2e1}^* = \frac{1}{A+B} \frac{2\Gamma_{\text{G},12} A - \Gamma_{\text{L},12} B}{\Gamma_{\text{L},11} + \Gamma_{\text{L},22}}.
\end{aligned} \tag{21a}$$

where

$$\begin{aligned}
A &= \Gamma_{\text{L},11}\Gamma_{\text{L},22} - \Gamma_{\text{L},12}\Gamma_{\text{L},21}, \\
B &= \Gamma_{\text{L},11}\Gamma_{\text{G},22} + \Gamma_{\text{L},22}\Gamma_{\text{G},11} - \Gamma_{\text{L},12}\Gamma_{\text{G},21} - \Gamma_{\text{L},21}\Gamma_{\text{G},12}.
\end{aligned} \tag{21b}$$

An interesting question is whether gain can be engineered to produce a steady state that breaks time-reversal symmetry. Time reversal invariance of the steady state density matrix requires that $\langle \mathcal{T} i | \rho_\text{S}(t=\infty) | \mathcal{T} j \rangle = \langle i | \rho_\text{S}(t=\infty) | j \rangle^*$ for generic atomic states

$|i\rangle,|j\rangle$. In particular, the steady-state has a broken time reversal symmetry when $\rho_{\mathrm{e1e1}} \neq \rho_{\mathrm{e2e2}}$. However, within the quasi-static approximation and assuming isotropic gain, this is not feasible. In this regime, the interaction tensors $\overline{\overline{\mathcal{G}}}_{\alpha}^{\mathrm{NH}}$ are real-valued [see Eq. (18)], which implies that the decay (or excitation) rates from (or into) the two excited states are identical, i.e., $\Gamma_{\mathrm{L},11} = \Gamma_{\mathrm{L},22}$ ($\Gamma_{\mathrm{G},11} = \Gamma_{\mathrm{G},22}$). As a result, from Eq. (21), the steady-state populations satisfy $\rho_{\mathrm{e1e1}} = \rho_{\mathrm{e2e2}}$, and time-reversal symmetry is preserved.

Furthermore, breaking time-reversal symmetry is only feasible when the transition dipole moment is elliptically (or circularly) polarized. In contrast, for linearly polarized dipole moments, it is straightforward to verify—regardless of the gain environment—that $\Gamma_{\mathrm{L},11} = \Gamma_{\mathrm{L},22} = |\Gamma_{\mathrm{L},12}| \equiv \Gamma_{\mathrm{L}}$ and $\Gamma_{\mathrm{G},11} = \Gamma_{\mathrm{G},22} = |\Gamma_{\mathrm{G},12}| \equiv \Gamma_{\mathrm{G}}$. In this case, the coefficients $A$ and $B$ in Eq. (21b) vanish, and consequently the steady-state given by Eq. (21a) becomes ill-defined. The reason is that for linear polarization, the null space of the Lindblad superoperator has dimension two, implying that the steady-state density matrix is not unique. A detailed analysis shows that it takes the form:

$$\begin{aligned}
\rho_{\mathrm{gg}} &= \frac{\Gamma_{\mathrm{L}}\sqrt{2}\cos\left(\theta-\frac{\pi}{4}\right)}{\Gamma_{\mathrm{L}}\sqrt{2}\cos\left(\theta-\frac{\pi}{4}\right)+2\Gamma_{\mathrm{G}}\cos\theta}, \\
\rho_{\mathrm{e1e1}} = \rho_{\mathrm{e2e2}} &= \frac{\Gamma_{\mathrm{G}}\cos\theta}{\Gamma_{\mathrm{L}}\sqrt{2}\cos\left(\theta-\frac{\pi}{4}\right)+2\Gamma_{\mathrm{G}}\cos\theta}, \\
\rho_{\mathrm{e1e2}} &= \frac{\Gamma_{\mathrm{G}}\sin\theta}{\Gamma_{\mathrm{L}}\sqrt{2}\cos\left(\theta-\frac{\pi}{4}\right)+2\Gamma_{\mathrm{G}}\cos\theta}.
\end{aligned} \tag{22}$$

where $-\pi/4 \leq \theta \leq \pi/2$ is an arbitrary real-valued coefficient. Thus, unlike the two-level system analyzed in Sec. IV.A, a qutrit with a V-shaped energy structure and linearly polarized transitions exhibits a steady state that depends on the initial condition. As a result, information about the initial state is not completely erased by the irreversible time evolution.

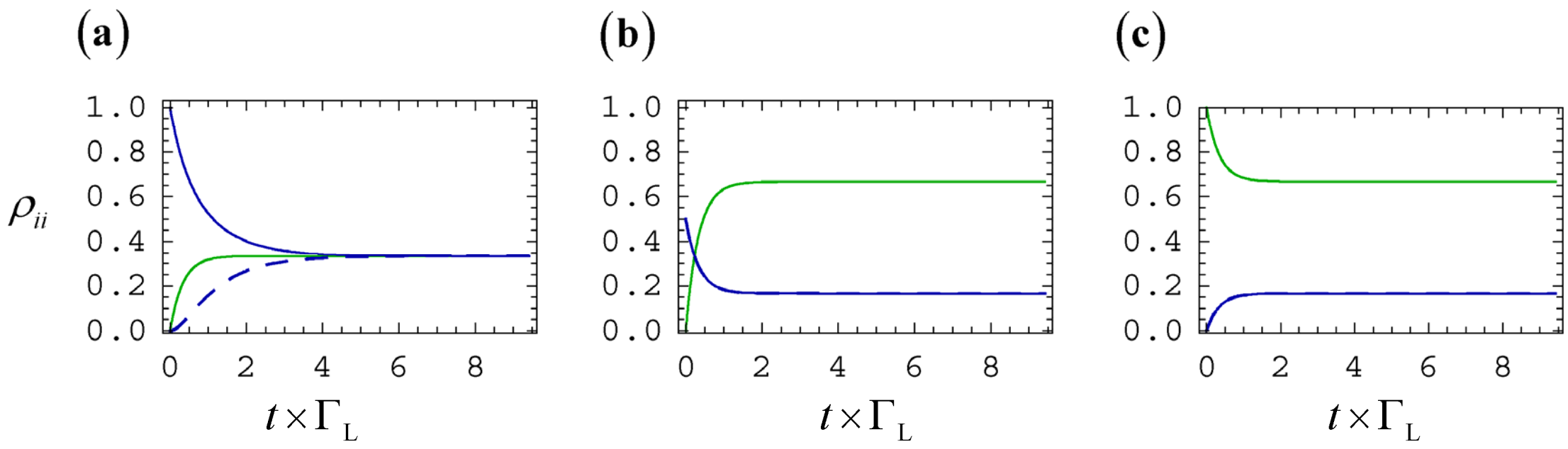

**Fig. 2** Time evolution of the atomic populations as a function of time for a qubit with a V-shaped energy structure and atomic transitions with linear polarization. Green line: $\rho_{gg}$. Solid blue line: $\rho_{e1e1}$. Dashed blue line: $\rho_{e2e2}$. In the simulations the decay and excitation rates are $\Gamma_L = 0.1\omega_a$ and $\Gamma_G = 0.5\Gamma_L$. The initial state is the pure state (a) $|e_1\rangle$, (b) $\frac{1}{\sqrt{2}}(|e_1\rangle + |e_2\rangle)$, and (c) $|g\rangle$.

This property is illustrated in Fig. 2, which shows the time evolution of the qutrit population in the zero-temperature limit for a linearly polarized transition. Clearly, the final-state populations can vary significantly depending on the initial conditions, highlighting the memory of the initial state retained in the steady state.

*C. Chiral-gain with physical motion*

In the final example, we continue to consider a qutrit with a V-shaped energy structure, as in the previous subsection, but now assume it interacts with a moving plasmonic substrate. The substrate moves at velocity *v* along the +*x* direction. Recent studies, particularly in the context of quantum friction [44-45, 61-64], have shown that physical motion can give rise to electromagnetic interactions with gain [20, 21]. In such systems, the non-Hermitian interactions, whether dissipative or amplifying, depend on the momentum of the electromagnetic waves relative to the direction of motion. Specifically, highly confined SPPs propagating along $+x$ with sufficiently short wavelengths may experience gain, while those propagating in the opposite direction invariably experience damping [20, 21]. Since the light–matter interaction is mediated by SPPs, which possess intrinsic chiral properties rooted in transverse spin [65-67], one may expect that the gain and dissipation in this system could exhibit a chiral character.

To investigate this, we computed the tensors that characterize the non-Hermitian interactions in this system, using a quasi-static approximation and considering weak (infinitesimal) material dissipation. As shown in Ref. [20], the quasi-static approximation accurately captures the interaction between a moving slab and an atom in the non-relativistic regime (e.g., $v < 0.1c$). The substrate is modeled using a standard Drude dispersion with a characteristic SPP resonance frequency $\omega_{\mathrm{sp}}$. In the non-relativistic limit, we find that the interaction tensors $\overline{\boldsymbol{\mathcal{G}}}_{\alpha}^{\mathrm{NH}}$ satisfy:

$$\overline{\boldsymbol{\mathcal{G}}}_{\alpha}^{\mathrm{NH}} \approx G_{\alpha,0}\frac{1}{2}\left(\hat{\mathbf{x}}+is_{\alpha}\hat{\mathbf{z}}\right)\otimes\left(\hat{\mathbf{x}}-is_{\alpha}\hat{\mathbf{z}}\right), \quad \text{with} \quad G_{\alpha,0}=\frac{1}{8\pi\varepsilon_0}k_{\alpha}^{2}\frac{\omega_{\mathrm{sp}}}{v}\sqrt{\frac{\pi}{\left|k_{\alpha}\right|z_{\mathrm{a}}}}e^{-2\left|k_{\alpha}\right|z_{\mathrm{a}}} \tag{23}$$

with $k_{\mathrm{L}}=\dfrac{\omega_{\mathrm{a}}-\omega_{\mathrm{sp}}}{v}$, $k_{\mathrm{G}}=\dfrac{\omega_{\mathrm{a}}+\omega_{\mathrm{sp}}}{v}$, $s_{\alpha}=\operatorname{sgn}k_{\alpha}$, and $\alpha=\mathrm{L,G}$. These expressions assume $2\left|k_{\alpha}\right|z_{\mathrm{a}} >> 1$, which is typically satisfied for non-relativistic velocities. Importantly, the only nontrivial eigenvector of $\overline{\boldsymbol{\mathcal{G}}}_{\alpha}^{\mathrm{NH}}$ corresponds to a circular polarization, $\left(\hat{\mathbf{x}}+is_{\alpha}\hat{\mathbf{z}}\right)/\sqrt{2}$. This property confirms that the non-Hermitian interactions in this system are inherently chiral, as expected from the fact that they are mediated by SPPs. In practical terms, this means that an atomic transition whose polarization matches the eigenvector of $\overline{\mathcal{G}}_{\mathrm{L}}^{\mathrm{NH}}$ ($\overline{\mathcal{G}}_{\mathrm{G}}^{\mathrm{NH}}$) will be associated with a non-conservative decay (excitation) process.

Moreover, provided $\omega_{\mathrm{a}} < \omega_{\mathrm{sp}}$ (which will be assumed in the following) the handedness of the eigenvector associated with $\overline{\mathcal{G}}_{\mathrm{L}}^{\mathrm{NH}}$ is opposite to the handedness of the eigenvector associated with $\overline{\mathcal{G}}_{\mathrm{G}}^{\mathrm{NH}}$ because $s_{\mathrm{L}} = -1 = -s_{\mathrm{G}}$.

To illustrate these ideas, let us suppose that the electric-dipole transition $|e_1\rangle \to |g\rangle$ is of the form $\boldsymbol{\gamma}_{\mathrm{e}} \approx \gamma_{\mathrm{e}} \frac{1}{\sqrt{2}}(\hat{\mathbf{x}} + i\hat{\mathbf{z}})$ so that it matches the handedness of $\overline{\mathcal{G}}_{\mathrm{G}}^{\mathrm{NH}}$. Then, by time-reversal symmetry, the dipole moment associated with the partner states $|e_2\rangle \to |g\rangle$ is $\boldsymbol{\gamma}_{\mathrm{e}}^* \approx \gamma_{\mathrm{e}} \frac{1}{\sqrt{2}}(\hat{\mathbf{x}} - i\hat{\mathbf{z}})$, which matches the handedness of $\overline{\mathcal{G}}_{\mathrm{L}}^{\mathrm{NH}}$. Thus, intuitively, the dissipative interactions promote the decay of the atomic state $|e_2\rangle$ to the ground, whereas the gain interactions promote the excitation of the ground to $|e_1\rangle$. Consistent with this heuristic picture, it can be readily shown that in the zero-temperature limit the relevant decay and excitation rates are

$$\Gamma_{22,\mathrm{L}} = \frac{2\gamma_{\mathrm{e}}^2 G_{\mathrm{L},0}}{\hbar}, \qquad \Gamma_{11,\mathrm{G}} = \frac{2\gamma_{\mathrm{e}}^2 G_{\mathrm{G},0}}{\hbar}. \tag{24}$$

with all other rates vanishing. Substituting these expressions into Eq. (21), one finds that the steady-state density matrix is $\hat{\rho}_{\mathrm{S},\infty} = |e_1\rangle\langle e_1|$, indicating that the qutrit always relaxes to the excited state associated with the gain transition. This corresponds to explicit time-reversal symmetry breaking, as $\rho_{\mathrm{e1e1}} = 1$ and $\rho_{\mathrm{e2e2}} = 0$. Hence, chiral gain, when combined with atomic transitions, allows for the selection of a skewed equilibrium state that violates time-reversal invariance. This mechanism leads to a nonreciprocal polarizability response and may also induce a nontrivial spin magnetic moment in the equilibrium state.

Although the analysis in this article focuses on moving platforms, more practical implementations of a chiral-gain response rely on either optical pumping of low-symmetry materials using chiral light [16], or on electrical biasing of low-symmetry conductors with a nontrivial quantum geometry [17, 18, 19].

The quasi-static approximation derived in Appendix E assumes infinitesimal dissipation in the metal and neglects the coupling to radiative modes (photons). While these effects can be rigorously accounted for using the full Green's function of the system, next we incorporate them in the analysis in a simplified manner by adding a scalar term to the interaction tensor $\overline{\mathcal{G}}_{\mathrm{L}}^{\mathrm{NH}}$, so that $\overline{\mathcal{G}}_{\mathrm{L}}^{\mathrm{NH}} \to \overline{\mathcal{G}}_{\mathrm{L}}^{\mathrm{NH}} + G_{00}\mathbf{1}_{3\times 3}$. Here, $G_{00}$ is intended to model the effect of stronger dissipation and the coupling with radiative modes. For example, the standard spontaneous emission process in free space corresponds to the term $G_{00} = \frac{1}{6\pi\varepsilon_0}\left(\frac{\omega_{\mathrm{a}}}{c}\right)^3$. With this correction, the excitation rates remain unchanged (namely, $\Gamma_{11,\mathrm{G}}$ remains the only nonzero excitation rate), whereas the nontrivial decay rates are now:

$$\Gamma_{11,\mathrm{L}} = \frac{2\gamma_{\mathrm{e}}^2 G_{00}}{\hbar}, \qquad \Gamma_{22,\mathrm{L}} = \frac{2\gamma_{\mathrm{e}}^2 G_{\mathrm{L},0}}{\hbar} + \frac{2\gamma_{\mathrm{e}}^2 G_{00}}{\hbar}. \tag{25}$$

In this case, the zero-temperature steady state becomes:

$$\hat{\rho}_{\mathrm{S},\infty} = \frac{\Gamma_{\mathrm{L},11}}{\Gamma_{\mathrm{L},11}+\Gamma_{\mathrm{G},11}}|g\rangle\langle g| + \frac{\Gamma_{\mathrm{G},11}}{\Gamma_{\mathrm{L},11}+\Gamma_{\mathrm{G},11}}|e_1\rangle\langle e_1|, \tag{26}$$

which still exhibits broken-time reversal symmetry. The steady state now depends on the relative strength between the loss and gain processes. Time-reversal symmetry breaking is more robust when the excitation rate $\Gamma_{\mathrm{G},11}$ significantly exceeds the decay rate $\Gamma_{\mathrm{L},11}$ arising from intrinsic dissipative mechanisms, such as electron collisions or spontaneous emission due to coupling with vacuum modes. Figure 3a presents a numerical example illustrating how the population of the excited state with chirality opposite to that which induces gain approaches zero over time.

We note in passing that the chiral transition associated with the state $|e_1\rangle$ is the one responsible for enhancing the ground-state quantum friction force [68] (here, the handedness is opposite to that considered in Ref. [68], since the relative velocity between the atom and the substrate differs by a minus sign).

The previous considerations neglect the temperature effect. In general, thermal fluctuations tend to suppress time-reversal symmetry breaking, as thermal agitation promotes upward and downward transitions equally, regardless of the atomic polarization. This effect is illustrated in Fig. 3b, which shows that at sufficiently high (effective) temperatures, the atomic population tends to equilibrate such that $\rho_{\mathrm{gg}} \approx \rho_{\mathrm{e1e1}} \approx \rho_{\mathrm{e2e2}} \approx 1/3$. Note that the horizontal axis of Fig. 3b represents the occupation number $N_\omega = \dfrac{1}{e^{+\beta\hbar\omega} - 1}$ evaluated at the atomic transition frequency, which is a function of the inverse (effective) temperature $\beta$ of the system. For temperatures such that $\hbar\omega_{\mathrm{a}} << k_B T$, we have $N_{\omega_{\mathrm{a}}} \approx \dfrac{k_B T}{\hbar\omega_{\mathrm{a}}}$, whereas for low-temperatures $k_B T << \hbar\omega_{\mathrm{a}}$, one has $N_{\omega_{\mathrm{a}}} \approx 0$.

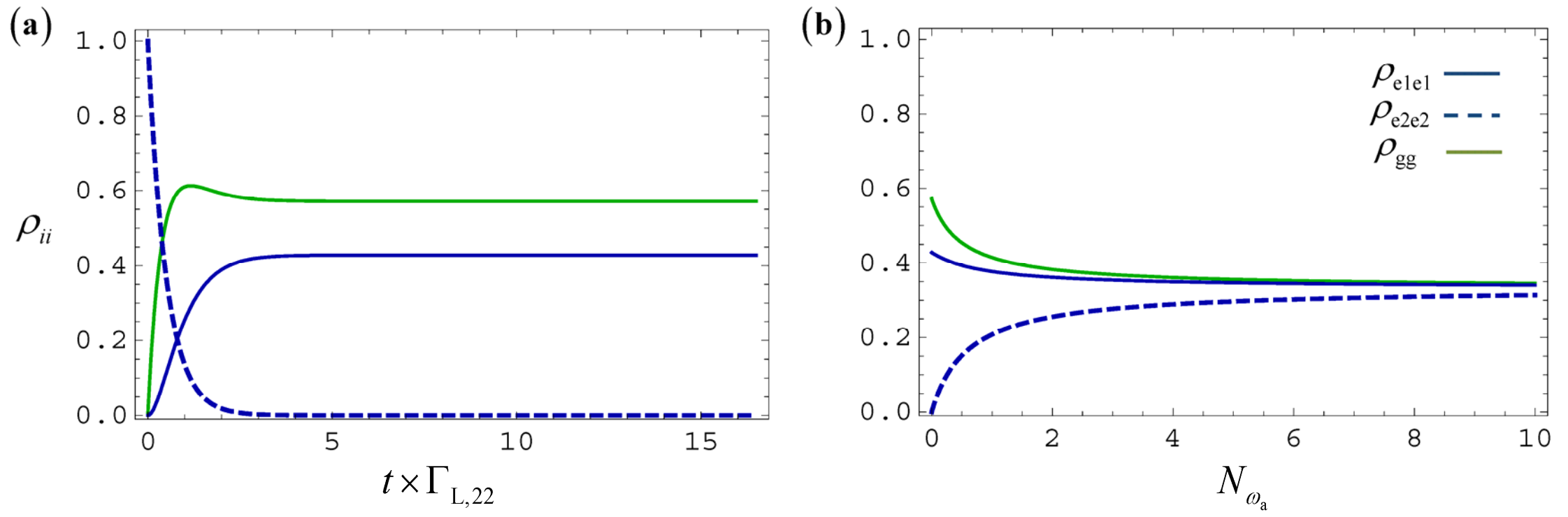

**Fig. 3** (a) Time evolution of the atomic populations as a function of time for a qutrit with a V-shaped energy structure and atomic transitions with circular polarization. The initial state is the pure state $|e_2\rangle$ and the zero-temperature limit is assumed. The parameters of the simulation are $\Gamma_{L,11} = 0.1\omega_a$, $\Gamma_{G,11} = 0.75\Gamma_{L,11}$ and $\Gamma_{L,22} = \Gamma_{L,11} + \Gamma_{G,11}$. (b) Steady-state atomic populations as a function of the occupation number $N_{\omega_a}$ (governed by the temperature) for the same parameters as in (a). Note that the decay and excitation rates at a finite temperature can be expressed in terms of the same rates at zero temperature (see Eq. (13)). In both plots the color code is the same as in the inset of panel b).

## V. Summary

In summary, we have extended the phenomenological quantization framework to encompass general macroscopic bianisotropic photonic systems exhibiting structured gain. Our approach assumes a causal and stable material response, ensuring that the photonic Green's function has its poles in the lower half of the complex frequency plane. We have emphasized that the decomposition of the response into gain and dissipation components depends on the underlying microscopic mechanisms governing the system's non-Hermitian dynamics.

Moreover, we have derived generalized fluctuation-dissipation relations for environments with structured gain described by a generalized Gibbs state. Building on this, we formulated a quantum master equation in Lindblad form that captures the time evolution of an atom embedded in gain environment. Notably, we demonstrated that the interplay between chiral atomic transitions and chiral gain can lead to an equilibrium atomic state with broken time-reversal symmetry, giving rise to nonreciprocal atomic behavior.

As a concrete example, we analyzed the interaction of a qutrit with a V-shaped energy level structure coupled to a moving plasmonic platform, explicitly showing that motion-induced gain can exhibit chiral properties in such a system. In future work, we aim to explore more practical platforms for realizing chiral gain, particularly those based on electro-optic effects in low-symmetry conductors [69]. We believe that our approach opens new avenues for tailoring light–matter interactions and engineering nonreciprocity in the quantum regime.

**Acknowledgements:** This work is supported in part by the Institution of Engineering and Technology (IET), by the Simons Foundation (award SFI-MPS-EWP-00008530-10), and by FCT/MECI through national funds and when applicable co-funded EU funds under UID/50008: Instituto de Telecomunicações. D.O. acknowledges the supports by JSPS Overseas Research Fellowship and by the RIKEN special postdoctoral researcher program.

## Appendix A: Quantized electromagnetic field in gain systems

Here, we extend the phenomenological theory of electromagnetic field quantization [22-24] to generic bi-anisotropic gain systems.

### *A. Classical system*

We consider a generic linear bianisotropic system whose electromagnetic response is described by a 6×6 material matrix $\mathbf{M} = \mathbf{M}(\mathbf{r},\omega)$ that determines the material response through the constitutive relations in Eq. (1).

The frequency-domain Maxwell's equations with electric ($\mathbf{j}_e$) and magnetic current sources ($\mathbf{j}_m$) are:

$$i\nabla\times\mathbf{H} - \omega\mathbf{D} = i\mathbf{j}_e, \qquad -i\nabla\times\mathbf{E} - \omega\mathbf{B} = i\mathbf{j}_m. \tag{A1}$$

The system Green's function $\overline{\mathcal{G}}(\mathbf{r},\mathbf{r}',\omega)$ (6×6 tensor) is the classical solution of the Maxwell's equation for dipole-type electric and magnetic current excitations:

$$\hat{\mathcal{D}}\cdot\overline{\mathcal{G}} - \omega\mathbf{M}(\mathbf{r},\omega)\cdot\overline{\mathcal{G}} = \omega\mathbf{1}\delta(\mathbf{r}-\mathbf{r}'), \tag{A2}$$

In the above $\mathbf{r}$ is the observation point, $\mathbf{r}'$ is the source point and $\hat{\mathcal{D}}$ is the differential operator

$$\hat{\mathcal{D}} = \begin{pmatrix} \mathbf{0}_{3\times3} & i\nabla\times\mathbf{1}_{3\times3} \\ -i\nabla\times\mathbf{1}_{3\times3} & \mathbf{0}_{3\times3} \end{pmatrix}. \tag{A3}$$

We suppose that the material matrix satisfies $\mathbf{M}(\omega) = \mathbf{M}^*(-\omega^*)$ so that the Green's function satisfies $\overline{\mathcal{G}}(\mathbf{r},\mathbf{r}',\omega) = \left[\overline{\mathcal{G}}(\mathbf{r},\mathbf{r}',-\omega)\right]^*$, in agreement with the reality symmetry of the electromagnetic field. Furthermore, the classical system is assumed stable so that all the normal modes are associated with an exponential decay in time, even in the presence of material gain. Thereby, the Green's function is analytical in the upper-half frequency plane. In addition, the material response for large frequencies should be coincident with the response of the vacuum, so that:

$$\lim_{\omega\to\infty}\mathbf{M}(\omega)=\mathbf{M}_0\equiv\begin{pmatrix}\varepsilon_0\mathbf{1}_{3\times3} & \mathbf{0}_{3\times3}\\ \mathbf{0}_{3\times3} & \mu_0\mathbf{1}_{3\times3}\end{pmatrix}. \tag{A4}$$

It is convenient to introduce a Green's function $\overline{\mathcal{G}}_{\rm c}(\mathbf{r},\mathbf{r}',\omega)$ for the Hermitian conjugate problem, defined such that $\left[\hat{\mathcal{D}}-\omega\mathbf{M}^\dagger(\mathbf{r},\omega)\right]\cdot\overline{\mathcal{G}}_{\rm c}=\omega\mathbf{1}\delta(\mathbf{r}-\mathbf{r}')$. Multiplying both members of this equation by $\overline{\mathcal{G}}^\dagger(\mathbf{r},\mathbf{r}'',\omega)$ and integrating by parts over $\mathbf{r}$ it is found that for real-valued $\omega$

$$\int d^3\mathbf{r}\left[\left[\hat{\mathcal{D}}-\omega\mathbf{M}(\mathbf{r},\omega)\right]\cdot\overline{\mathcal{G}}(\mathbf{r},\mathbf{r}'',\omega)\right]^\dagger\cdot\overline{\mathcal{G}}_{\rm c}(\mathbf{r},\mathbf{r}',\omega)=\omega\overline{\mathcal{G}}^\dagger(\mathbf{r}',\mathbf{r}'',\omega). \tag{A5}$$

Hence, from Eq. (A2) it follows that the two Green's functions are related as $\overline{\mathcal{G}}_{\rm c}(\mathbf{r},\mathbf{r}',\omega)=\overline{\mathcal{G}}^\dagger(\mathbf{r}',\mathbf{r},\omega)$. Evidently, we have that

$$\left[\hat{\mathcal{D}}-\omega\mathbf{M}(\mathbf{r},\omega)\right]\cdot\left[\overline{\mathcal{G}}-\overline{\mathcal{G}}_{\rm c}\right]=\omega\left[\mathbf{M}(\mathbf{r},\omega)-\mathbf{M}^\dagger(\mathbf{r},\omega)\right]\cdot\overline{\mathcal{G}}_{\rm c} \tag{A6}$$

Therefore, it is clear that $\overline{\mathcal{G}}-\overline{\mathcal{G}}_{\rm c}=\overline{\mathcal{G}}(\mathbf{r},\mathbf{r}',\omega)-\left[\overline{\mathcal{G}}(\mathbf{r}',\mathbf{r},\omega)\right]^\dagger$ has the following Green's function representation:

$$\overline{\mathcal{G}}(\mathbf{r},\mathbf{r}',\omega)-\left[\overline{\mathcal{G}}(\mathbf{r}',\mathbf{r},\omega)\right]^\dagger=\int d^3\mathbf{r}_0\;\overline{\mathcal{G}}(\mathbf{r},\mathbf{r}_0,\omega)\cdot\left[\mathbf{M}(\mathbf{r}_0,\omega)-\mathbf{M}^\dagger(\mathbf{r}_0,\omega)\right]\cdot\left[\overline{\mathcal{G}}(\mathbf{r}',\mathbf{r}_0,\omega)\right]^\dagger \tag{A7}$$

The above identity holds true for arbitrary real-valued $\omega$.

*B. Quantized fields*

Next, we extend the standard phenomenological quantization of the electromagnetic fields in macroscopic systems to the case of gain media [31, 32]. To this end, we introduce two sets of bosonic operators $\hat{\mathbf{b}}_{{\rm L},\omega}(\mathbf{r}_0)$ and $\hat{\mathbf{b}}_{{\rm G},\omega}(\mathbf{r}_0)$ satisfying canonical commutation relations:

$$\left[\hat{\mathbf{b}}_{\alpha,\omega}\left(\mathbf{r}_0\right),\hat{\mathbf{b}}^{\dagger}_{\alpha',\omega'}\left(\mathbf{r}'_0\right)\right]=\mathbf{1}_{6\times6}\delta\left(\omega-\omega'\right)\delta\left(\mathbf{r}_0-\mathbf{r}'_0\right)\delta_{\alpha\alpha'}, \tag{A8a}$$

$$\left[\hat{\mathbf{b}}_{\alpha,\omega}\left(\mathbf{r}_0\right),\hat{\mathbf{b}}_{\alpha',\omega'}\left(\mathbf{r}'_0\right)\right]=0=\left[\hat{\mathbf{b}}^{\dagger}_{\alpha,\omega}\left(\mathbf{r}_0\right),\hat{\mathbf{b}}^{\dagger}_{\alpha',\omega'}\left(\mathbf{r}'_0\right)\right], \tag{A8b}$$

with $\alpha=\mathrm{L,G}$ and $\alpha'=\mathrm{L,G}$. By definition, the element *m,n* of the left-hand side tensor in Eq. (A8a) is $\left[\hat{\mathbf{u}}_m\cdot\hat{\mathbf{b}}_{\alpha,\omega}\left(\mathbf{r}_0\right),\hat{\mathbf{b}}^{\dagger}_{\alpha',\omega'}\left(\mathbf{r}'_0\right)\cdot\hat{\mathbf{u}}_n\right]$. Note that all the operators with index L commute with all the operators with index G. Here, $\hat{\mathbf{u}}_m$ a 6-dimensional unit vector with 1 in the *m*-th position and 0s elsewhere.

The bosonic operators $\hat{\mathbf{b}}_{\mathrm{L},\omega}\left(\mathbf{r}_0\right)$ are associated with the dissipative part of the system response, whereas the bosonic operators $\hat{\mathbf{b}}_{\mathrm{G},\omega}\left(\mathbf{r}_0\right)$ are associated with the gain part of the response. The Hamiltonian of the quantized system is written in terms of the bosonic operators as:

$$\hat{H}_{\mathrm{EM}}=\int_0^{\infty}d\omega\int d^3\mathbf{r}\,\hbar\omega\left[\hat{\mathbf{b}}^{\dagger}_{\mathrm{L},\omega}\left(\mathbf{r}\right)\cdot\hat{\mathbf{b}}_{\mathrm{L},\omega}\left(\mathbf{r}\right)-\hat{\mathbf{b}}^{\dagger}_{\mathrm{G},\omega}\left(\mathbf{r}\right)\cdot\hat{\mathbf{b}}_{\mathrm{G},\omega}\left(\mathbf{r}\right)\right]. \tag{A9}$$

Evidently, for a free-field theory, the time-variation of the bosonic operators in the Heisenberg picture is:

$$\hat{\mathbf{b}}_{\mathrm{L},\omega}\left(\mathbf{r},t\right)=\hat{\mathbf{b}}_{\mathrm{L},\omega}\left(\mathbf{r}\right)e^{-i\omega t},\qquad \hat{\mathbf{b}}_{\mathrm{G},\omega}\left(\mathbf{r},t\right)=\hat{\mathbf{b}}_{\mathrm{G},\omega}\left(\mathbf{r}\right)e^{+i\omega t}. \tag{A10}$$

The quantized (free) electromagnetic fields $\hat{\mathbf{E}}$ and $\hat{\mathbf{H}}$ are conveniently defined in terms of the 6-component vector $\hat{\mathbf{F}}=\left(\hat{\mathbf{E}}\quad\hat{\mathbf{H}}\right)^T$ and of the photonic Green's function as follows:

$$\hat{\mathbf{F}}=\hat{\mathbf{F}}_{\mathrm{L}}+\hat{\mathbf{F}}_{\mathrm{G}}+\left[\hat{\mathbf{F}}_{\mathrm{L}}+\hat{\mathbf{F}}_{\mathrm{G}}\right]^{\dagger},\qquad\text{with} \tag{A11a}$$

$$\hat{\mathbf{F}}_{\mathrm{L}}(\mathbf{r},t)=\int_0^{\infty} d\omega\, e^{-i\omega t}\sqrt{\frac{\hbar}{\pi|\omega|}}\int d^3\mathbf{r}_0\, \overline{\mathcal{G}}(\mathbf{r},\mathbf{r}_0,\omega)\cdot \mathbf{R}_{\mathrm{L}\omega}(\mathbf{r}_0)\cdot \hat{\mathbf{b}}_{\mathrm{L},\omega}(\mathbf{r}_0), \qquad \text{(A11b)}$$

$$\hat{\mathbf{F}}_{\mathrm{G}}(\mathbf{r},t)=\int_0^{\infty} d\omega\, e^{-i\omega t}\sqrt{\frac{\hbar}{\pi|\omega|}}\int d^3\mathbf{r}_0\, \overline{\mathcal{G}}(\mathbf{r},\mathbf{r}_0,\omega)\cdot \mathbf{R}_{\mathrm{G}\omega}(\mathbf{r}_0)\cdot \hat{\mathbf{b}}^{\dagger}_{\mathrm{G},\omega}(\mathbf{r}_0)\cdot \qquad \text{(A11c)}$$

We introduced the multiplication operators $\mathbf{R}_{\alpha,\omega}$ ($\alpha=\mathrm{L,G}$) determined by the loss and gain components of $\mathbf{M}''$ in Eq. (3):

$$\mathbf{R}_{\mathrm{L}\omega}=\left[\omega\mathbf{M}''_{\mathrm{L}}\right]^{1/2}, \qquad \mathbf{R}_{\mathrm{G}\omega}=\left[-\omega\mathbf{M}''_{\mathrm{G}}\right]^{1/2}. \qquad \text{(A12)}$$

It is implicit that $\omega>0$. The operators $\mathbf{R}_{\alpha,\omega}$ are non-negative Hermitian matrices. For future reference, we note that:

$$\mathbf{R}_{\mathrm{L}\omega}\cdot\mathbf{R}^{\dagger}_{\mathrm{L}\omega}=\omega\mathbf{M}''_{\mathrm{L}}, \qquad \mathbf{R}_{\mathrm{G}\omega}\cdot\mathbf{R}^{\dagger}_{\mathrm{G}\omega}=-\omega\mathbf{M}''_{\mathrm{G}}. \qquad \text{(A13)}$$

From the Green's function definition [Eq. (A2)], the equivalent noise current operators are:

$$\hat{\mathbf{j}}_{\mathrm{N}}(\mathbf{r}_0,t)=-i\int_0^{+\infty} d\omega\, e^{-i\omega t}\sqrt{\frac{\hbar|\omega|}{\pi}}\left[\mathbf{R}_{\mathrm{L}\omega}(\mathbf{r}_0)\cdot\hat{\mathbf{b}}_{\mathrm{L},\omega}(\mathbf{r}_0)+\mathbf{R}_{\mathrm{G}\omega}(\mathbf{r}_0)\cdot\hat{\mathbf{b}}^{\dagger}_{\mathrm{G},\omega}(\mathbf{r}_0)\right]+h.c. \qquad \text{(A14)}$$

Here, "h.c." stands for Hermitian conjugate.

In the following subsection, we confirm that the quantized fields satisfy canonical commutation equations.

*C. Commutation relations*

Next, we find the commutation relations for the different field components $\left[\hat{F}_i(\mathbf{r}),\hat{F}_j(\mathbf{r}')\right]\equiv\hat{F}_i(\mathbf{r})\hat{F}_j(\mathbf{r}')-\hat{F}_j(\mathbf{r}')\hat{F}_i(\mathbf{r})$ (*i,j*=1…6). From Eqs. (A8) and (A11), it is evident that:

$$\begin{aligned}\left[\hat{F}_i(\mathbf{r}),\hat{F}_j(\mathbf{r}')\right]=&\left[\hat{F}_{\mathrm{L},i}(\mathbf{r}),\hat{F}^\dagger_{\mathrm{L},j}(\mathbf{r}')\right]+\left[\hat{F}_{\mathrm{G},i}(\mathbf{r}),\hat{F}^\dagger_{\mathrm{G},j}(\mathbf{r}')\right]\\&-\left[\hat{F}_{\mathrm{L},j}(\mathbf{r}'),\hat{F}^\dagger_{\mathrm{L},i}(\mathbf{r})\right]-\left[\hat{F}_{\mathrm{G},j}(\mathbf{r}'),\hat{F}^\dagger_{\mathrm{G},i}(\mathbf{r})\right]\end{aligned}. \tag{A15}$$

A straightforward analysis shows that

$$\left[\hat{F}_{\mathrm{L},i}(\mathbf{r}),\hat{F}^\dagger_{\mathrm{L},j}(\mathbf{r}')\right]+\left[\hat{F}_{\mathrm{G},i}(\mathbf{r}),\hat{F}^\dagger_{\mathrm{G},j}(\mathbf{r}')\right]=$$
$$\int_0^\infty d\omega\int d^3\mathbf{r}_0\ \frac{\hbar}{\pi|\omega|}\hat{\mathbf{u}}_i\cdot\overline{\mathcal{G}}(\mathbf{r},\mathbf{r}_0,\omega)\cdot\left[\mathbf{R}_{\mathrm{L}\omega}(\mathbf{r}_0)\cdot\mathbf{R}^\dagger_{\mathrm{L}\omega}(\mathbf{r}_0)-\mathbf{R}_{\mathrm{G}\omega}(\mathbf{r}_0)\cdot\mathbf{R}^\dagger_{\mathrm{G}\omega}(\mathbf{r}_0)\right]\cdot\left[\overline{\mathcal{G}}(\mathbf{r}',\mathbf{r}_0,\omega)\right]^\dagger\cdot\hat{\mathbf{u}}_j \tag{A16}$$

Thus, with the help of Eqs. (3) and (A13), one can write the commutation relations as:

$$\begin{aligned}\left[\hat{\mathbf{F}}(\mathbf{r}),\hat{\mathbf{F}}(\mathbf{r}')\right]=&\ \frac{\hbar}{\pi}\int_0^\infty d\omega\int d^3\mathbf{r}_0\ \overline{\mathcal{G}}(\mathbf{r},\mathbf{r}_0,\omega)\cdot\mathbf{M}''(\mathbf{r}_0)\cdot\left[\overline{\mathcal{G}}(\mathbf{r}',\mathbf{r}_0,\omega)\right]^\dagger\\&-\mathrm{transpose}(\mathbf{r}\leftrightarrow\mathbf{r}')\end{aligned} \tag{A17}$$

By definition, the element $\hat{\mathbf{u}}_i\cdot\left[\hat{\mathbf{F}}(\mathbf{r}),\hat{\mathbf{F}}(\mathbf{r}')\right]\cdot\hat{\mathbf{u}}_j$ stands for $\left[\hat{F}_i(\mathbf{r}),\hat{F}_j(\mathbf{r}')\right]$. Next, we use the identity (A7) to perform the spatial integration explicitly:

$$\begin{aligned}\left[\hat{\mathbf{F}}(\mathbf{r}),\hat{\mathbf{F}}(\mathbf{r}')\right]=&\ \frac{\hbar}{2i\pi}\int_0^\infty d\omega\left[\overline{\mathcal{G}}(\mathbf{r},\mathbf{r}',\omega)-\left[\overline{\mathcal{G}}(\mathbf{r}',\mathbf{r},\omega)\right]^\dagger+c.c.\right]\\=&\ \frac{\hbar}{2i\pi}\int_{-\infty}^\infty d\omega\left[\overline{\mathcal{G}}(\mathbf{r},\mathbf{r}',\omega)-\left[\overline{\mathcal{G}}(\mathbf{r}',\mathbf{r},\omega)\right]^T\right]\end{aligned} \tag{A18}$$

Here, "c.c." stands for the complex conjugated term. In the second identity we used the reality property of the Green's function $\overline{\mathcal{G}}(\mathbf{r},\mathbf{r}',\omega)=\left[\overline{\mathcal{G}}(\mathbf{r},\mathbf{r}',-\omega)\right]^*$ to transform the integration domain into the entire real-frequency axis.

As the classical system is stable, the Green's function is analytical in the upper-half frequency plane. Thus, the integral over the real-line can be deformed into an integral over a semicircle with infinite radius in the upper-half frequency plane. Taking into account the asymptotic behavior of the material matrix for large frequency [Eq. (A4)], it follows that for $\omega\to\infty$ the Green's function satisfies $\overline{\mathcal{G}}(\mathbf{r},\mathbf{r}',\omega)\approx\overline{\mathcal{G}}_0(\mathbf{r},\mathbf{r}',\omega)$, where $\overline{\mathcal{G}}_0(\mathbf{r},\mathbf{r}',\omega)$ is the free-space Green's function. This proves that the commutation relations can only depend on $\overline{\mathcal{G}}_0(\mathbf{r},\mathbf{r}',\omega)$, and thereby must be same as for the vacuum. The vacuum commutation relations, which as we have just argued also hold true for the dispersive gain material system, can be written in a compact form as [31]:

$$\left[\mathbf{M}_0\cdot\hat{\mathbf{F}}(\mathbf{r}),\mathbf{M}_0\cdot\hat{\mathbf{F}}(\mathbf{r}')\right]=\hbar\hat{\mathcal{D}}\cdot\mathbf{1}_{6\times6}\,\delta(\mathbf{r}-\mathbf{r}'),\tag{A19}$$

where $\hat{\mathcal{D}}$ is defined as in Eq. (A3) and $\mathbf{M}_0$ is the material matrix of the vacuum [Eq. (A4)].

Phenomenological quantization often relies on introducing a transverse vector potential in the Coulomb gauge [32]. A similar construction can be readily formulated within our framework. Specifically, we can introduce a transverse six-vector potential operator $\hat{\mathbf{A}}=\begin{pmatrix}\hat{\mathbf{A}}_1 & \hat{\mathbf{A}}_2\end{pmatrix}^T$ given by $\hat{\mathbf{A}}=\hat{\mathbf{A}}_\mathrm{L}+\hat{\mathbf{A}}_\mathrm{G}+\left[\hat{\mathbf{A}}_\mathrm{L}+\hat{\mathbf{A}}_\mathrm{G}\right]^\dagger$, where $\hat{\mathbf{A}}_\mathrm{L},\hat{\mathbf{A}}_\mathrm{G}$ are defined as:

$$\hat{\mathbf{A}}_\mathrm{L}(\mathbf{r},t)=\boldsymbol{\delta}^\perp_{6\times6}*\int_0^\infty d\omega\,e^{-i\omega t}\frac{1}{i\omega}\sqrt{\frac{\hbar}{\pi|\omega|}}\int d^3\mathbf{r}_0\,\overline{\mathcal{G}}(\mathbf{r},\mathbf{r}_0,\omega)\cdot\mathbf{R}_{\mathrm{L}\omega}(\mathbf{r}_0)\cdot\hat{\mathbf{b}}_{\mathrm{L},\omega}(\mathbf{r}_0),\tag{A20a}$$

$$\hat{\mathbf{A}}_\mathrm{G}(\mathbf{r},t)=\boldsymbol{\delta}^\perp_{6\times6}*\int_0^\infty d\omega\,e^{-i\omega t}\frac{1}{i\omega}\sqrt{\frac{\hbar}{\pi|\omega|}}\int d^3\mathbf{r}_0\,\overline{\mathcal{G}}(\mathbf{r},\mathbf{r}_0,\omega)\cdot\mathbf{R}_{\mathrm{G}\omega}(\mathbf{r}_0)\cdot\hat{\mathbf{b}}^\dagger_{\mathrm{G},\omega}(\mathbf{r}_0).\tag{A20b}$$

In the above, "*" represents a spatial convolution, $\boldsymbol{\delta}^\perp_{6\times6}=\begin{pmatrix}\boldsymbol{\delta}^\perp & 0\\ 0 & \boldsymbol{\delta}^\perp\end{pmatrix}$ and $\boldsymbol{\delta}^\perp$ stands for the standard transverse delta-function operator. A straightforward analysis shows that

$$\left[\hat{\mathbf{F}}(\mathbf{r}),\hat{\mathbf{A}}(\mathbf{r}')\right]=\hbar\mathbf{M}_0^{-1}\cdot\boldsymbol{\delta}^\perp_{6\times6}(\mathbf{r}-\mathbf{r}'),\tag{A21}$$

which corresponds to the standard canonical commutation relations between the field and the vector potential.

It is relevant to note that in our approach the electric and magnetic fields are treated on equal footing. Moreover, unlike other methods (see [32]), for systems with a nontrivial magnetic response, our noise currents can exhibit both electric and magnetic components. As a result, our commutation relations maintain a high degree of symmetry between the two fields, with each commutation relation for the electric field (e.g., $\left[\hat{\mathbf{E}}(\mathbf{r}),\hat{\mathbf{E}}(\mathbf{r}')\right]=0$) having a corresponding counterpart for the magnetic field (e.g., $\left[\hat{\mathbf{H}}(\mathbf{r}),\hat{\mathbf{H}}(\mathbf{r}')\right]=0$).

## Appendix B: Quantized field correlations

In this Appendix, we derive the quantized field correlations. For clarity, we first calculate these for the vacuum state, which corresponds to the environment's density matrix $\hat{\rho}_{\mathrm{E}}=|0_{\mathrm{E}}\rangle\langle 0_{\mathrm{E}}|$, and then extend the results to the generalized Gibbs state. Finally, we utilize the derived expressions to obtain a generalized fluctuation-dissipation relation.

*A. Field correlations for the vacuum state*

In the following, we demonstrate that for $\hat{\rho}_{\mathrm{E}}=|0_{\mathrm{E}}\rangle\langle 0_{\mathrm{E}}|$ the quantized fields satisfy:

$$\mathrm{tr}\left\{\hat{\rho}_{\mathrm{E}}\hat{\mathbf{F}}(\mathbf{r},t)\otimes\hat{\mathbf{F}}(\mathbf{r}',t-\tau)\right\}=\frac{\hbar}{2\pi i}\int_0^{\infty}d\omega\left[e^{-i\omega\tau}\mathbf{\Lambda}^{\mathrm{L}}_{\mathbf{r},\mathbf{r}'}(\omega)-e^{i\omega\tau}\mathbf{\Lambda}^{\mathrm{G},*}_{\mathbf{r},\mathbf{r}'}(\omega)\right],\tag{B1a}$$

$$\mathrm{tr}\left\{\hat{\mathbf{F}}(\mathbf{r},t)\hat{\rho}_{\mathrm{E}}\otimes\hat{\mathbf{F}}(\mathbf{r}',t-\tau)\right\}=-\frac{\hbar}{2\pi i}\int_0^{\infty}d\omega\left[e^{+i\omega\tau}\mathbf{\Lambda}^{\mathrm{L},*}_{\mathbf{r},\mathbf{r}'}(\omega)-e^{-i\omega\tau}\mathbf{\Lambda}^{\mathrm{G}}_{\mathbf{r},\mathbf{r}'}(\omega)\right],\tag{B1b}$$

with

$$\mathbf{\Lambda}^{\mathrm{L}}_{\mathbf{r},\mathbf{r}'}(\omega)=+2i\int d^3\mathbf{r}_0\,\overline{\mathcal{G}}(\mathbf{r},\mathbf{r}_0,\omega)\cdot\mathbf{M}''_{\mathrm{L}}(\mathbf{r}_0)\cdot\overline{\mathcal{G}}^{\dagger}(\mathbf{r}',\mathbf{r}_0,\omega),\tag{B2a}$$

$$\mathbf{\Lambda}^{\mathrm{G}}_{\mathbf{r},\mathbf{r}'}(\omega) = -2i\int d^3\mathbf{r}_0\, \overline{\mathcal{G}}(\mathbf{r},\mathbf{r}_0,\omega)\cdot \mathbf{M}''_{\mathrm{G}}(\mathbf{r}_0)\cdot \overline{\mathcal{G}}^{\dagger}(\mathbf{r}',\mathbf{r}_0,\omega). \tag{B2b}$$

We start with Eq. (B1a). With the help of Eq. (A11), one readily finds that

$$\begin{aligned}\mathrm{tr}\left\{\hat{\rho}_{\mathrm{E}}\hat{\mathbf{F}}(\mathbf{r},t)\otimes\hat{\mathbf{F}}(\mathbf{r}',t-\tau)\right\} = {} & \frac{\hbar}{\pi}\int_0^{\infty} d\omega e^{-i\omega\tau}\int d^3\mathbf{r}_0\, \overline{\mathcal{G}}(\mathbf{r},\mathbf{r}_0,\omega)\cdot\mathbf{M}''_{\mathrm{L}}(\mathbf{r}_0)\cdot\overline{\mathcal{G}}(\mathbf{r}',\mathbf{r}_0,\omega)^{\dagger} \\ & -\frac{\hbar}{\pi}\left[\int_0^{\infty} d\omega e^{-i\omega\tau}\int d^3\mathbf{r}_0\, \overline{\mathcal{G}}(\mathbf{r},\mathbf{r}_0,\omega)\cdot\mathbf{M}''_{\mathrm{G}}(\mathbf{r}_0)\cdot\overline{\mathcal{G}}(\mathbf{r}',\mathbf{r}_0,\omega)^{\dagger}\right]^{*}\end{aligned} \tag{B3}$$

We used $\hat{\rho}_{\mathrm{E}} = \left|0_{\mathrm{E}}\right\rangle\left\langle 0_{\mathrm{E}}\right|$, the bosonic commutation relations [Eq. (A8)] and $\hat{\mathbf{b}}_{\alpha,\omega}\left|0_{\mathrm{E}}\right\rangle = 0$, with $\alpha = \mathrm{L,G}$, and Eq. (A13). It is evident that Eq. (B3) can be expressed as in Eq. (B1a), as we wanted to show.

In order to demonstrate Eq. (B1b), we note that due to the cyclic property of the trace:

$$\mathrm{tr}\left\{\hat{\mathbf{F}}_i(\mathbf{r},t)\,\hat{\rho}_{\mathrm{E}}\hat{\mathbf{F}}_j(\mathbf{r}',t-\tau)\right\} = \mathrm{tr}\left\{\hat{\rho}_{\mathrm{E}}\hat{\mathbf{F}}_j(\mathbf{r}',t-\tau)\,\hat{\mathbf{F}}_i(\mathbf{r},t)\right\} = \mathrm{tr}\left\{\hat{\rho}_{\mathrm{E}}\hat{\mathbf{F}}_j(\mathbf{r}',t)\,\hat{\mathbf{F}}_i(\mathbf{r},t+\tau)\right\}. \tag{B4}$$

In the last identity, we used the fact that the field-field correlations are stationary in time. Then, from Eq. (B1a) one finds that:

$$\mathrm{tr}\left\{\hat{\mathbf{F}}(\mathbf{r},t)\,\hat{\rho}_{\mathrm{E}}\otimes\hat{\mathbf{F}}(\mathbf{r}',t-\tau)\right\} = \frac{\hbar}{2\pi i}\int_0^{\infty} d\omega\left[e^{+i\omega\tau}\mathbf{\Lambda}^{\mathrm{L}}_{\mathbf{r}',\mathbf{r}}(\omega) - e^{-i\omega\tau}\mathbf{\Lambda}^{\mathrm{G},*}_{\mathbf{r}',\mathbf{r}}(\omega)\right]^{T}. \tag{B5}$$

The superscript $T$ stands for the tensor transposition. Using the identity $\left[\mathbf{\Lambda}^{\mathrm{L/G}}_{\mathbf{r}',\mathbf{r}}(\omega)\right]^{T} = -\left[\mathbf{\Lambda}^{\mathrm{L/G}}_{\mathbf{r},\mathbf{r}'}(\omega)\right]^{*}$ one obtains Eq. (B1b), which concludes the proof.

*B. Field correlations for the generalized Gibbs state*

Next, we extend Eq. (B1a) to a generalized Gibbs state defined as in Eq. (7) with $\beta_{\mathrm{L}} = -\beta_{\mathrm{G}} \equiv \beta$. It is evident that:

$$\mathrm{tr}\left\{\hat{\rho}_{\mathrm{E}}\hat{\mathbf{F}}(\mathbf{r},t)\otimes\hat{\mathbf{F}}(\mathbf{r}',t-\tau)\right\}=\mathrm{tr}\left\{\hat{\rho}_{\mathrm{E}}\hat{\mathbf{F}}_{\mathrm{L}}^{\dagger}(\mathbf{r},t)\otimes\hat{\mathbf{F}}_{\mathrm{L}}(\mathbf{r}',t-\tau)\right\}+\mathrm{tr}\left\{\hat{\rho}_{\mathrm{E}}\hat{\mathbf{F}}_{\mathrm{L}}(\mathbf{r},t)\otimes\hat{\mathbf{F}}_{\mathrm{L}}^{\dagger}(\mathbf{r}',t-\tau)\right\}$$
$$+\mathrm{tr}\left\{\hat{\rho}_{\mathrm{E}}\hat{\mathbf{F}}_{\mathrm{G}}^{\dagger}(\mathbf{r},t)\otimes\hat{\mathbf{F}}_{\mathrm{G}}(\mathbf{r}',t-\tau)\right\}+\mathrm{tr}\left\{\hat{\rho}_{\mathrm{E}}\hat{\mathbf{F}}_{\mathrm{G}}(\mathbf{r},t)\otimes\hat{\mathbf{F}}_{\mathrm{G}}^{\dagger}(\mathbf{r}',t-\tau)\right\} \quad \text{(B6)}$$

As the bosonic operators satisfy canonical commutation relations [Eq. (A8)] and as the operator $\beta_{\mathrm{L}}\hat{H}_{\mathrm{EM,L}}+\beta_{\mathrm{G}}\hat{H}_{\mathrm{EM,G}}$ in definition of the generalized Gibbs state (7) has the standard positive definite structure when $\beta_{\mathrm{L}}=-\beta_{\mathrm{G}}\equiv\beta>0$, one has:

$$\mathrm{tr}\left\{\hat{\rho}_{\mathrm{E}}\hat{\mathbf{b}}_{\alpha,\omega}^{\dagger}(\mathbf{r}_0)\otimes\hat{\mathbf{b}}_{\alpha,\omega'}(\mathbf{r}_0')\right\}=\mathbf{1}_{6\times6}\,\delta(\omega-\omega')\,\delta(\mathbf{r}_0-\mathbf{r}_0')\,N_{\omega}\,. \quad \text{(B7a)}$$

$$\mathrm{tr}\left\{\hat{\rho}_{\mathrm{E}}\hat{\mathbf{b}}_{\alpha,\omega}(\mathbf{r}_0)\otimes\hat{\mathbf{b}}_{\alpha,\omega'}^{\dagger}(\mathbf{r}_0')\right\}=\mathbf{1}_{6\times6}\,\delta(\omega-\omega')\,\delta(\mathbf{r}_0-\mathbf{r}_0')\,(N_{\omega}+1)\,. \quad \text{(B7b)}$$

where $N_{\omega}=\frac{1}{e^{+\beta\hbar\omega}-1}$ is the expectation of the occupation number operator for an harmonic oscillator at (effective) temperature $T$ and $\alpha=\mathrm{L,G}$. Using the above formulas and Eq. (A11) one can readily evaluate each of the terms in Eq. (B6):

$$\mathrm{tr}\left\{\hat{\rho}_{\mathrm{E}}\hat{\mathbf{F}}_{\mathrm{L}}(\mathbf{r},t)\otimes\hat{\mathbf{F}}_{\mathrm{L}}^{\dagger}(\mathbf{r}',t-\tau)\right\}=\frac{\hbar}{\pi}\int_0^{\infty}d\omega\,e^{-i\omega\tau}(1+N_{\omega})\int d^3\mathbf{r}_0\,\overline{\boldsymbol{\mathcal{G}}}(\mathbf{r},\mathbf{r}_0,\omega)\cdot\mathbf{M}_{\mathrm{L}}''(\mathbf{r}_0)\cdot\overline{\boldsymbol{\mathcal{G}}}(\mathbf{r}',\mathbf{r}_0,\omega)^{\dagger} \quad \text{(B8a)}$$

$$\mathrm{tr}\left\{\hat{\rho}_{\mathrm{E}}\hat{\mathbf{F}}_{\mathrm{L}}^{\dagger}(\mathbf{r},t)\otimes\hat{\mathbf{F}}_{\mathrm{L}}(\mathbf{r}',t-\tau)\right\}=\frac{\hbar}{\pi}\left[\int_0^{\infty}d\omega\,e^{-i\omega\tau}N_{\omega}\int d^3\mathbf{r}_0\,\overline{\boldsymbol{\mathcal{G}}}(\mathbf{r},\mathbf{r}_0,\omega)\cdot\mathbf{M}_{\mathrm{L}}''(\mathbf{r}_0)\cdot\overline{\boldsymbol{\mathcal{G}}}(\mathbf{r}',\mathbf{r}_0,\omega)^{\dagger}\right]^{*} \quad \text{(B8b)}$$

$$\mathrm{tr}\left\{\hat{\rho}_{\mathrm{E}}\hat{\mathbf{F}}_{\mathrm{G}}(\mathbf{r},t)\otimes\hat{\mathbf{F}}_{\mathrm{G}}^{\dagger}(\mathbf{r}',t-\tau)\right\}=-\frac{\hbar}{\pi}\int_0^{\infty}d\omega\,e^{-i\omega\tau}N_{\omega}\int d^3\mathbf{r}_0\,\overline{\boldsymbol{\mathcal{G}}}(\mathbf{r},\mathbf{r}_0,\omega)\cdot\mathbf{M}_{\mathrm{G}}''(\mathbf{r}_0)\cdot\overline{\boldsymbol{\mathcal{G}}}(\mathbf{r}',\mathbf{r}_0,\omega)^{\dagger} \quad \text{(B8c)}$$

$$\mathrm{tr}\left\{\hat{\rho}_{\mathrm{E}}\hat{\mathbf{F}}_{\mathrm{G}}^{\dagger}\left(\mathbf{r},t\right)\otimes\hat{\mathbf{F}}_{\mathrm{G}}\left(\mathbf{r}',t-\tau\right)\right\}$$

$$=-\frac{\hbar}{\pi}\left[\int_0^\infty d\omega e^{-i\omega\tau}\left(1+N_\omega\right)\int d^3\mathbf{r}_0\,\overline{\mathcal{G}}\left(\mathbf{r},\mathbf{r}_0,\omega\right)\cdot\mathbf{M}''_{\mathrm{G}}\left(\mathbf{r}_0\right)\cdot\overline{\mathcal{G}}\left(\mathbf{r}',\mathbf{r}_0,\omega\right)^\dagger\right]^*$$

(B8d)

Substituting these results in Eq. (B6), one can show that the field correlations for the generalized Gibbs state are determined by:

$$\begin{aligned}\mathrm{tr}\left\{\hat{\rho}_{\mathrm{E}}\hat{\mathbf{F}}\left(\mathbf{r},t\right)\otimes\hat{\mathbf{F}}\left(\mathbf{r}',t-\tau\right)\right\}=&\\ &\frac{\hbar}{\pi}\int_0^\infty d\omega e^{-i\omega\tau}\int d^3\mathbf{r}_0\,\overline{\mathcal{G}}\left(\mathbf{r},\mathbf{r}_0,\omega\right)\cdot\mathbf{M}''_{\mathrm{L}}\left(\mathbf{r}_0\right)\cdot\overline{\mathcal{G}}\left(\mathbf{r}',\mathbf{r}_0,\omega\right)^\dagger\\ &\frac{\hbar}{\pi}2\,\mathrm{Re}\int_0^\infty d\omega e^{-i\omega\tau}N_\omega\int d^3\mathbf{r}_0\,\overline{\mathcal{G}}\left(\mathbf{r},\mathbf{r}_0,\omega\right)\cdot\mathbf{M}''_{\mathrm{L}}\left(\mathbf{r}_0\right)\cdot\overline{\mathcal{G}}\left(\mathbf{r}',\mathbf{r}_0,\omega\right)^\dagger\\ &-\frac{\hbar}{\pi}\left[\int_0^\infty d\omega e^{-i\omega\tau}\int d^3\mathbf{r}_0\,\overline{\mathcal{G}}\left(\mathbf{r},\mathbf{r}_0,\omega\right)\cdot\mathbf{M}''_{\mathrm{G}}\left(\mathbf{r}_0\right)\cdot\overline{\mathcal{G}}\left(\mathbf{r}',\mathbf{r}_0,\omega\right)^\dagger\right]^*\\ &-\frac{\hbar}{\pi}2\,\mathrm{Re}\int_0^\infty d\omega e^{-i\omega\tau}N_\omega\int d^3\mathbf{r}_0\,\overline{\mathcal{G}}\left(\mathbf{r},\mathbf{r}_0,\omega\right)\cdot\mathbf{M}''_{\mathrm{G}}\left(\mathbf{r}_0\right)\cdot\overline{\mathcal{G}}\left(\mathbf{r}',\mathbf{r}_0,\omega\right)^\dagger\end{aligned} \quad \text{(B9)}$$

In the zero temperature limit, $N_\omega \to 0$ and the above result reduces to Eq. (B3), as it should be.

*C. Fluctuation-dissipation relation for the free-fields*

Next, we derive a generalized fluctuation-dissipation relation for gain systems. To this end, we denote $\left\langle\left\{\hat{\mathbf{F}}\left(\mathbf{r},t\right)\hat{\mathbf{F}}\left(\mathbf{r}',t\right)\right\}_+\right\rangle_T$ as the tensor obtained from the Gibbs state expectation of the symmetrized field products:

$$\hat{\mathbf{u}}_i\cdot\left\langle\left\{\hat{\mathbf{F}}\left(\mathbf{r},t\right)\hat{\mathbf{F}}\left(\mathbf{r}',t\right)\right\}_+\right\rangle_T\cdot\hat{\mathbf{u}}_j\equiv\frac{1}{2}\mathrm{tr}\left\{\hat{\rho}_{\mathrm{E}}\left(\hat{F}_i\left(\mathbf{r},t\right)\hat{F}_j\left(\mathbf{r}',t\right)+\hat{F}_j\left(\mathbf{r}',t\right)\hat{F}_i\left(\mathbf{r},t\right)\right)\right\},\quad i,j=1\ldots6$$

(B10)

It is clear that:

$$\left\langle \left\{ \hat{\mathbf{F}}(\mathbf{r},t)\hat{\mathbf{F}}(\mathbf{r}',t) \right\}_+ \right\rangle_T = \frac{1}{2}\mathrm{tr}\left\{ \hat{\rho}_{\mathrm{E}} \hat{\mathbf{F}}(\mathbf{r},t) \otimes \hat{\mathbf{F}}(\mathbf{r}',t) \right\} + \left( \frac{1}{2}\mathrm{tr}\left\{ \hat{\rho}_{\mathrm{E}} \hat{\mathbf{F}}(\mathbf{r}',t) \otimes \hat{\mathbf{F}}(\mathbf{r},t) \right\} \right)^T . \quad \text{(B11)}$$

From Eq. (B9), it follows that:

$$\left\langle \left\{ \hat{\mathbf{F}}(\mathbf{r},t)\hat{\mathbf{F}}(\mathbf{r}',t) \right\}_+ \right\rangle_T = \int_0^\infty d\omega \left\langle \left\{ \hat{\mathbf{F}}(\mathbf{r})\hat{\mathbf{F}}(\mathbf{r}') \right\}_+ \right\rangle_{T,\omega}, \qquad \text{with} \qquad \text{(B12a)}$$

$$\begin{aligned} &\left\langle \left\{ \hat{\mathbf{F}}(\mathbf{r})\hat{\mathbf{F}}(\mathbf{r}') \right\}_+ \right\rangle_{T,\omega} \\ &= \frac{2\hbar}{\pi}\left( N_\omega + \frac{1}{2} \right) \mathrm{Re} \int d^3\mathbf{r}_0 \, \overline{\mathcal{G}}(\mathbf{r},\mathbf{r}_0,\omega) \cdot \left[ \mathbf{M}''_{\mathrm{L}}(\mathbf{r}_0) - \mathbf{M}''_{\mathrm{G}}(\mathbf{r}_0) \right] \cdot \overline{\mathcal{G}}(\mathbf{r}',\mathbf{r}_0,\omega)^\dagger \end{aligned} \qquad \text{(B12b)}$$

As discussed in the main text, this relation generalizes the standard fluctuation dissipation relation in passive systems [41-42]. Note that $N_\omega + \frac{1}{2} = \frac{1}{2}\coth\left( \frac{\beta\hbar\omega}{2} \right)$.

Similarly, it is possible to show that the noise current [Eq. (A14)] correlations are determined by:

$$\left\langle \left\{ \hat{\mathbf{j}}_{\mathrm{N}}(\mathbf{r},t)\hat{\mathbf{j}}_{\mathrm{N}}(\mathbf{r}',t) \right\}_+ \right\rangle_T = \delta(\mathbf{r}-\mathbf{r}')\mathrm{Re}\frac{2\hbar}{\pi}\int_0^\infty d\omega \left( N_\omega + \frac{1}{2} \right) \omega^2 \left[ \mathbf{M}''_{\mathrm{L}}(\mathbf{r},\omega) - \mathbf{M}''_{\mathrm{G}}(\mathbf{r},\omega) \right].$$

(B13)

## Appendix C: Derivation of the quantum master equation

### *A. Generic quantum system*

To begin with, we consider a quantum system described by the Hamiltonian $\hat{H}_{\mathrm{at}}$. The quantum system is formed by a set of atoms ($n = 1,2,...$), with the *n*-th atom positioned at the point $\mathbf{r}_n$. The coupling of the atoms with the environment is modeled by the interaction Hamiltonian

$$\hat{H}_{\mathrm{int}} = -\sum_n \hat{\mathbf{F}}(\mathbf{r}_n) \cdot \hat{\mathbf{p}}_n , \qquad \text{(C1)}$$

where $\mathbf{r}_n$ represents the atom's coordinates and $\hat{\mathbf{p}}_n = \begin{pmatrix} \hat{\mathbf{p}}_{\mathrm{e}}^{(n)} & \hat{\mathbf{p}}_{\mathrm{m}}^{(n)} \end{pmatrix}^T$ is the (6-vector) transition dipole moment operator of the *n*-th atom. The sub-component $\hat{\mathbf{p}}_{\mathrm{e}}$ is the usual electric dipole moment operator, whereas $\hat{\mathbf{p}}_{\mathrm{m}} = \mu_0 \hat{\mathbf{m}}_{\mathrm{s}}$ is determined by the (spin) magnetic dipole moment operator ($\hat{\mathbf{m}}_{\mathrm{s}}$), which for generality is included in the formulation.

We employ the standard density matrix formalism to describe the time-evolution of the quantum system. The full-system density matrix is denoted by $\hat{\rho}$. It satisfies the von Neumann equation $\partial_t \hat{\rho} = \frac{i}{\hbar}\left[\hat{\rho}, \hat{H}\right]$ with $\hat{H} = \hat{H}_0 + \hat{H}_{\mathrm{int}}$ and $\hat{H}_0 = \hat{H}_{\mathrm{at}} + \hat{H}_{\mathrm{EM}}$. The time evolution of the density matrix in the interaction picture, $\hat{\rho}_{\mathrm{I}} = e^{+i\frac{\hat{H}_0}{\hbar}t} \hat{\rho} e^{-i\frac{\hat{H}_0}{\hbar}t}$, is determined by:

$$\partial_t \hat{\rho}_{\mathrm{I}} = \frac{i}{\hbar}\left[\hat{\rho}_{\mathrm{I}}, \hat{H}_{\mathrm{int}}(t)\right], \tag{C2}$$

with $\hat{H}_{\mathrm{int}}(t) = e^{i\frac{\hat{H}_0}{\hbar}t} \hat{H}_{\mathrm{int}} e^{-i\frac{\hat{H}_0}{\hbar}t}$.

In the Born approximation, the density matrix is assumed to satisfy $\hat{\rho}_{\mathrm{I}}(t) \approx \hat{\rho}_{\mathrm{S,I}}(t) \otimes \hat{\rho}_{\mathrm{E}}$ at all time instants, where $\hat{\rho}_{\mathrm{S,I}}$ and $\hat{\rho}_{\mathrm{E}}$ depend only on the atomic (environment) states, respectively. For simplicity in this Appendix, we take $\hat{\rho}_{\mathrm{E}} \approx |0_{\mathrm{E}}\rangle\langle 0_{\mathrm{E}}|$ with $|0_{\mathrm{E}}\rangle$ the environment state with no excitations. The effect of a finite temperature (modeled by $\hat{\rho}_{\mathrm{E}} \approx \hat{\rho}_{\mathrm{E,Gibbs}}$) is analyzed in Appendix D.

Proceeding in a standard manner [47], one may show that the dynamics of the reduced density matrix $\hat{\rho}_{\mathrm{S,I}}(t)$ is governed by

$$\partial_t \hat{\rho}_{\mathrm{S,I}} = -\frac{1}{\hbar^2}\int_0^{+\infty} \mathrm{tr}_{\mathrm{E}}\left\{\left[\left[\hat{\rho}_{\mathrm{I}}(t-\tau), \hat{H}_{\mathrm{int}}(t-\tau)\right], \hat{H}_{\mathrm{int}}(t)\right]\right\} d\tau . \tag{C3}$$

The operator $\mathrm{tr}_{\mathrm{E}}\{...\}$ represents the trace over the environment degrees of freedom. Using the Markov approximation, $\hat{\rho}_{\mathrm{I}}(t-\tau) \to \hat{\rho}_{\mathrm{I}}(t)$ and $\hat{H}_{\mathrm{int}}(t) = -\sum_n \hat{\mathbf{F}}(\mathbf{r}_n, t)\cdot \hat{\mathbf{p}}_n(t)$, we obtain:

$$\partial_t \hat{\rho}_{\mathrm{S,I}} = \mathcal{L}\hat{\rho}_{\mathrm{S,I}}, \qquad \text{with } \mathcal{L}\hat{\rho}_{\mathrm{S,I}} = \mathrm{I+II+III+IV} \tag{C4}$$

and

$$\mathrm{I} = -\frac{1}{\hbar^2}\hat{\rho}_{\mathrm{S,I}}(t)\sum_{m,n}\int_0^{+\infty} d\tau\, \hat{\mathbf{p}}_m(t-\tau)\cdot \mathrm{tr}_{\mathrm{E}}\left\{\hat{\rho}_{\mathrm{E}}\hat{\mathbf{F}}(\mathbf{r}_m, t-\tau)\otimes \hat{\mathbf{F}}(\mathbf{r}_n, t)\right\}\cdot \hat{\mathbf{p}}_n(t),$$

$$\mathrm{II} = \frac{1}{\hbar^2}\sum_{m,n}\int_0^{+\infty} d\tau\, \hat{\mathbf{p}}_m(t-\tau)\cdot \mathrm{tr}_{\mathrm{E}}\left\{\hat{\mathbf{F}}(\mathbf{r}_m, t-\tau)\hat{\rho}_{\mathrm{E}}\otimes \hat{\mathbf{F}}(\mathbf{r}_n, t)\right\}\hat{\rho}_{\mathrm{S,I}}(t)\cdot \hat{\mathbf{p}}_n(t),$$

$$\mathrm{III} = \frac{1}{\hbar^2}\sum_{m,n}\int_0^{+\infty} d\tau\, \hat{\mathbf{p}}_m(t)\cdot \mathrm{tr}_{\mathrm{E}}\left\{\hat{\mathbf{F}}(\mathbf{r}_m, t)\hat{\rho}_{\mathrm{E}}\otimes \hat{\mathbf{F}}(\mathbf{r}_n, t-\tau)\right\}\hat{\rho}_{\mathrm{S,I}}(t)\cdot \hat{\mathbf{p}}_n(t-\tau),$$

$$\mathrm{IV} = -\frac{1}{\hbar^2}\sum_{m,n}\int_0^{+\infty} d\tau\, \hat{\mathbf{p}}_m(t)\cdot \mathrm{tr}_{\mathrm{E}}\left\{\hat{\mathbf{F}}(\mathbf{r}_m, t)\otimes \hat{\mathbf{F}}(\mathbf{r}_n, t-\tau)\hat{\rho}_{\mathrm{E}}\right\}\cdot \hat{\mathbf{p}}_n(t-\tau)\hat{\rho}_{\mathrm{S,I}}(t).$$

The electromagnetic-field correlations are explicitly evaluated in Appendix B. Substituting Eq. (B1) into Eq. (C4), it is found that the Lindblad operator takes the form:

$$\begin{aligned}
&\mathcal{L}\hat{\rho}_{\mathrm{S,I}} = \\
&-\frac{1}{2\pi i}\frac{1}{\hbar}\hat{\rho}_{\mathrm{S,I}}(t)\sum_{m,n}\int_0^{\infty} d\omega \int_0^{+\infty} d\tau\, \hat{\mathbf{p}}_m(t-\tau)\cdot\left[e^{i\omega\tau}\mathbf{\Lambda}_{mn}^{\mathrm{L}}(\omega) - e^{-i\omega\tau}\mathbf{\Lambda}_{mn}^{\mathrm{G},*}(\omega)\right]\cdot \hat{\mathbf{p}}_n(t) \\
&-\frac{1}{2\pi i}\frac{1}{\hbar}\sum_{m,n}\int_0^{\infty} d\omega \int_0^{+\infty} d\tau\, \hat{\mathbf{p}}_m(t)\cdot\left[e^{-i\omega\tau}\mathbf{\Lambda}_{mn}^{\mathrm{L}}(\omega) - e^{+i\omega\tau}\mathbf{\Lambda}_{mn}^{\mathrm{G},*}(\omega)\right]\cdot \hat{\mathbf{p}}_n(t-\tau)\hat{\rho}_{\mathrm{S,I}}(t) \\
&-\frac{1}{2\pi i}\frac{1}{\hbar}\sum_{m,n}\int_0^{\infty} d\omega \int_0^{+\infty} d\tau\, \hat{\mathbf{p}}_m(t-\tau)\cdot\left[e^{-i\omega\tau}\mathbf{\Lambda}_{mn}^{\mathrm{L},*}(\omega) - e^{+i\omega\tau}\mathbf{\Lambda}_{mn}^{\mathrm{G}}(\omega)\right]\hat{\rho}_{\mathrm{S,I}}(t)\cdot \hat{\mathbf{p}}_n(t) \\
&-\frac{1}{2\pi i}\frac{1}{\hbar}\sum_{m,n}\int_0^{\infty} d\omega \int_0^{+\infty} d\tau\, \hat{\mathbf{p}}_m(t)\cdot\left[e^{+i\omega\tau}\mathbf{\Lambda}_{mn}^{\mathrm{L},*}(\omega) - e^{-i\omega\tau}\mathbf{\Lambda}_{mn}^{\mathrm{G}}(\omega)\right]\hat{\rho}_{\mathrm{S,I}}(t)\cdot \hat{\mathbf{p}}_n(t-\tau),
\end{aligned} \tag{C5}$$

with

$$\mathbf{\Lambda}_{mn}^{\mathrm{L/G}}(\omega) = \pm 2i\int d^3\mathbf{r}_0\, \overline{\mathcal{G}}(\mathbf{r}_m,\mathbf{r}_0,\omega)\cdot \mathbf{M}''_{\mathrm{L/G}}(\mathbf{r}_0)\cdot \overline{\mathcal{G}}^{\dagger}(\mathbf{r}_n,\mathbf{r}_0,\omega). \tag{C6}$$

For passive systems, it is useful to note that $\mathbf{\Lambda}_{mn}^{\mathrm{G}} = 0$ and $\mathbf{\Lambda}_{mn}^{\mathrm{L}}(\omega) = \overline{\mathcal{G}}(\mathbf{r}_m \mid \mathbf{r}_n,\omega) - \left[\overline{\mathcal{G}}(\mathbf{r}_m \mid \mathbf{r}_n,\omega)\right]^{\dagger}$, where we used Eq. (A7). In general, for systems with gain, the coefficients of the Lindblad operator must be evaluated numerically by integrating Eq. (C6).

*B. Degenerate two-level systems*

Next, we focus on system with a single atom with only two (potentially degenerate) energy levels, as detailed in Sect. III.A. Furthermore, we neglect the effects of the spin magnetic moment operator so that $\hat{\mathbf{p}} \approx \begin{pmatrix}\hat{\mathbf{p}}_{\mathrm{e}} & \mathbf{0}\end{pmatrix}^T$.

Within these hypotheses, the dipole operator may be assumed of the type $\hat{\mathbf{p}}(t) = \hat{\mathbf{p}}^-(t) + \hat{\mathbf{p}}^+(t)$ with $\hat{\mathbf{p}}^-(t) = \hat{\mathbf{p}}^- e^{-i\omega_{\mathrm{a}}t}$, $\hat{\mathbf{p}}^+(t) = \hat{\mathbf{p}}^+ e^{+i\omega_{\mathrm{a}}^* t}$ and $\hat{\mathbf{p}}^{\pm}$ ($\hat{\mathbf{p}}^+ = \left[\hat{\mathbf{p}}^-\right]^{\dagger}$) the operators in the Schrödinger picture determined only by the electric dipole operator. The operators $\hat{\mathbf{p}}^{\pm}$ take the form $\hat{\mathbf{p}}^{\pm} = \begin{pmatrix}\hat{\mathbf{p}}_{\mathrm{e}}^{\pm} & \mathbf{0}\end{pmatrix}^T$, with $\hat{\mathbf{p}}_{\mathrm{e}}^{\pm}$ defined as in Eq. (11). To account for a slow (adiabatic) switching of the interaction term, we include in $\omega_{\mathrm{a}}$ a small positive imaginary part ($\omega_{\mathrm{a}} \to \omega_{\mathrm{a}} + 0^+ i$).

Under the outlined conditions, the time delayed dipole operator can be written as $\hat{\mathbf{p}}(t-\tau) = \hat{\mathbf{p}}^-(t)e^{+i\omega_{\mathrm{a}}\tau} + \hat{\mathbf{p}}^+(t)e^{-i\omega_{\mathrm{a}}^*\tau}$, which allows us to evaluate explicitly the integrals in $\tau$ in Eq. (C5). Such procedure enables us to write $\mathcal{L}\hat{\rho}_{\mathrm{S,I}} = \mathcal{L}_{\mathrm{L}}\hat{\rho}_{\mathrm{S,I}} + \mathcal{L}_{\mathrm{G}}\hat{\rho}_{\mathrm{S,I}}$, with:

$$\mathcal{L}_{\mathrm{L}}\hat{\rho}_{\mathrm{S,I}} =$$
$$\begin{aligned}
&-\frac{i}{\hbar}\frac{1}{2\pi i}\hat{\rho}_{\mathrm{S,I}}(t)\int_{0^+}^{\infty} d\omega\left[\hat{\mathbf{p}}^-(t)\cdot\frac{\mathbf{\Lambda}_{\mathrm{L}}(\omega)}{\omega+\omega_{\mathrm{a}}}+\hat{\mathbf{p}}^+(t)\cdot\frac{\mathbf{\Lambda}_{\mathrm{L}}(\omega)}{\omega-\omega_{\mathrm{a}}^*}\right]\cdot\hat{\mathbf{p}}(t)\\
&+\frac{i}{\hbar}\frac{1}{2\pi i}\int_{0^+}^{\infty} d\omega\hat{\mathbf{p}}(t)\cdot\left[\frac{\mathbf{\Lambda}_{\mathrm{L}}(\omega)}{\omega-\omega_{\mathrm{a}}}\cdot\hat{\mathbf{p}}^-(t)+\frac{\mathbf{\Lambda}_{\mathrm{L}}(\omega)}{\omega+\omega_{\mathrm{a}}^*}\cdot\hat{\mathbf{p}}^+(t)\right]\hat{\rho}_{\mathrm{S,I}}(t)\\
&+\frac{i}{\hbar}\frac{1}{2\pi i}\int_{0^+}^{\infty} d\omega\left[\hat{\mathbf{p}}^-(t)\cdot\frac{\mathbf{\Lambda}_{\mathrm{L}}^*(\omega)}{\omega-\omega_{\mathrm{a}}}+\hat{\mathbf{p}}^+(t)\cdot\frac{\mathbf{\Lambda}_{\mathrm{L}}^*(\omega)}{\omega+\omega_{\mathrm{a}}^*}\right]\hat{\rho}_{\mathrm{S,I}}(t)\cdot\hat{\mathbf{p}}(t)\\
&-\frac{i}{\hbar}\frac{1}{2\pi i}\int_{0^+}^{\infty} d\omega\hat{\mathbf{p}}(t)\cdot\hat{\rho}_{\mathrm{S,I}}(t)\left[\frac{\mathbf{\Lambda}_{\mathrm{L}}^*(\omega)}{\omega+\omega_{\mathrm{a}}}\cdot\hat{\mathbf{p}}^-(t)+\frac{\mathbf{\Lambda}_{\mathrm{L}}^*(\omega)}{\omega-\omega_{\mathrm{a}}^*}\cdot\hat{\mathbf{p}}^+(t)\right],
\end{aligned} \tag{C7a}$$

$$\mathcal{L}_{\mathrm{G}}\hat{\rho}_{\mathrm{S,I}} =$$
$$\begin{aligned}
&-\frac{i}{\hbar}\frac{1}{2\pi i}\hat{\rho}_{\mathrm{S,I}}(t)\int_{0^+}^{\infty} d\omega\left[\hat{\mathbf{p}}^-(t)\cdot\frac{\mathbf{\Lambda}_{\mathrm{G}}^*(\omega)}{\omega-\omega_{\mathrm{a}}}+\hat{\mathbf{p}}^+(t)\cdot\frac{\mathbf{\Lambda}_{\mathrm{G}}^*(\omega)}{\omega+\omega_{\mathrm{a}}^*}\right]\cdot\hat{\mathbf{p}}(t)\\
&+\frac{i}{\hbar}\frac{1}{2\pi i}\int_{0^+}^{\infty} d\omega\hat{\mathbf{p}}(t)\cdot\left[\frac{\mathbf{\Lambda}_{\mathrm{G}}^*(\omega)}{\omega+\omega_{\mathrm{a}}}\cdot\hat{\mathbf{p}}^-(t)+\frac{\mathbf{\Lambda}_{\mathrm{G}}^*(\omega)}{\omega-\omega_{\mathrm{a}}^*}\cdot\hat{\mathbf{p}}^+(t)\right]\hat{\rho}_{\mathrm{S,I}}(t)\\
&+\frac{i}{\hbar}\frac{1}{2\pi i}\int_{0^+}^{\infty} d\omega\left[\hat{\mathbf{p}}^-(t)\cdot\frac{\mathbf{\Lambda}_{\mathrm{G}}(\omega)}{\omega+\omega_{\mathrm{a}}}+\hat{\mathbf{p}}^+(t)\cdot\frac{\mathbf{\Lambda}_{\mathrm{G}}(\omega)}{\omega-\omega_{\mathrm{a}}^*}\right]\hat{\rho}_{\mathrm{S,I}}(t)\cdot\hat{\mathbf{p}}(t)\\
&-\frac{i}{\hbar}\frac{1}{2\pi i}\int_{0^+}^{\infty} d\omega\hat{\mathbf{p}}(t)\cdot\hat{\rho}_{\mathrm{S,I}}(t)\left[\frac{\mathbf{\Lambda}_{\mathrm{G}}(\omega)}{\omega-\omega_{\mathrm{a}}}\cdot\hat{\mathbf{p}}^-(t)+\frac{\mathbf{\Lambda}_{\mathrm{G}}(\omega)}{\omega+\omega_{\mathrm{a}}^*}\cdot\hat{\mathbf{p}}^+(t)\right].
\end{aligned} \tag{C7b}$$

Here,

$$\mathbf{\Lambda}_{\mathrm{L/G}}(\omega) = \pm 2i\int d^3\mathbf{r}_0\, \overline{\boldsymbol{\mathcal{G}}}(\mathbf{r}_{\mathrm{a}},\mathbf{r}_0,\omega)\cdot\mathbf{M}''_{\mathrm{L/G}}(\mathbf{r}_0)\cdot\overline{\boldsymbol{\mathcal{G}}}^{\dagger}(\mathbf{r}_{\mathrm{a}},\mathbf{r}_0,\omega), \tag{C8}$$

with $\mathbf{r}_{\mathrm{a}}$ the atom position.

To proceed we introduce the auxiliary tensors:

$$\overline{\boldsymbol{\mathcal{G}}}_{\alpha}^{+} = \frac{1}{2\pi i}\int_{0^+}^{\infty} d\omega\frac{\mathbf{\Lambda}_{\alpha}(\omega)}{\omega-\omega_{\mathrm{a}}}, \qquad \overline{\boldsymbol{\mathcal{G}}}_{\alpha}^{-} = \frac{-1}{2\pi i}\int_{0^+}^{\infty} d\omega\frac{\mathbf{\Lambda}_{\alpha}^*(\omega)}{\omega+\omega_{\mathrm{a}}^*}, \qquad \alpha = \mathrm{L,G}. \tag{C9}$$

Taking into account that $\mathbf{\Lambda}_{\mathrm{L/G}}$ is an anti-Hermitian tensor ($\mathbf{\Lambda}_{\mathrm{L/G}}^{\dagger} = -\mathbf{\Lambda}_{\mathrm{L/G}}$), each of the terms of the Lindblad operator can be expressed as:

$$
\begin{aligned}
&\mathcal{L}_{\mathrm{L}}\hat{\rho}_{\mathrm{S,I}} = \\
&-\frac{i}{\hbar}\hat{\rho}_{\mathrm{S,I}}(t)\left[\hat{\mathbf{p}}^{-}(t)\cdot\left(\overline{\mathcal{G}}_{\mathrm{L}}^{-}\right)^{*}\cdot\hat{\mathbf{p}}(t)+\hat{\mathbf{p}}^{+}(t)\cdot\left(\overline{\mathcal{G}}_{\mathrm{L}}^{+}\right)^{\dagger}\cdot\hat{\mathbf{p}}(t)\right] \\
&+\frac{i}{\hbar}\left[\hat{\mathbf{p}}(t)\cdot\overline{\mathcal{G}}_{\mathrm{L}}^{+}\cdot\hat{\mathbf{p}}^{-}(t)+\hat{\mathbf{p}}(t)\cdot\left(\overline{\mathcal{G}}_{\mathrm{L}}^{-}\right)^{T}\cdot\hat{\mathbf{p}}^{+}(t)\right]\hat{\rho}_{\mathrm{S,I}}(t) \\
&-\frac{i}{\hbar}\left[\hat{\mathbf{p}}^{-}(t)\cdot\left(\overline{\mathcal{G}}_{\mathrm{L}}^{+}\right)^{T}+\hat{\mathbf{p}}^{+}(t)\cdot\overline{\mathcal{G}}_{\mathrm{L}}^{-}\right]\hat{\rho}_{\mathrm{S,I}}(t)\cdot\hat{\mathbf{p}}(t) \\
&+\frac{i}{\hbar}\hat{\mathbf{p}}(t)\cdot\hat{\rho}_{\mathrm{S,I}}(t)\left[\left(\overline{\mathcal{G}}_{\mathrm{L}}^{-}\right)^{\dagger}\cdot\hat{\mathbf{p}}^{-}(t)+\left(\overline{\mathcal{G}}_{\mathrm{L}}^{+}\right)^{*}\cdot\hat{\mathbf{p}}^{+}(t)\right],
\end{aligned}
\tag{C10a}
$$

$$
\begin{aligned}
&\mathcal{L}_{\mathrm{G}}\hat{\rho}_{\mathrm{S,I}} = \\
&+\frac{i}{\hbar}\hat{\rho}_{\mathrm{S,I}}(t)\left[\hat{\mathbf{p}}^{-}(t)\cdot\left(\overline{\mathcal{G}}_{\mathrm{G}}^{+}\right)^{T}\cdot\hat{\mathbf{p}}(t)+\hat{\mathbf{p}}^{+}(t)\cdot\overline{\mathcal{G}}_{\mathrm{G}}^{-}\cdot\hat{\mathbf{p}}(t)\right] \\
&-\frac{i}{\hbar}\left[\hat{\mathbf{p}}(t)\cdot\left(\overline{\mathcal{G}}_{\mathrm{G}}^{-}\right)^{\dagger}\cdot\hat{\mathbf{p}}^{-}(t)+\hat{\mathbf{p}}(t)\cdot\left(\overline{\mathcal{G}}_{\mathrm{G}}^{+}\right)^{*}\cdot\hat{\mathbf{p}}^{+}(t)\right]\hat{\rho}_{\mathrm{S,I}}(t) \\
&+\frac{i}{\hbar}\left[\hat{\mathbf{p}}^{-}(t)\cdot\left(\overline{\mathcal{G}}_{\mathrm{G}}^{-}\right)^{*}+\hat{\mathbf{p}}^{+}(t)\cdot\left(\overline{\mathcal{G}}_{\mathrm{G}}^{+}\right)^{\dagger}\right]\hat{\rho}_{\mathrm{S,I}}(t)\cdot\hat{\mathbf{p}}(t) \\
&-\frac{i}{\hbar}\hat{\mathbf{p}}(t)\cdot\hat{\rho}_{\mathrm{S,I}}(t)\left[\overline{\mathcal{G}}_{\mathrm{G}}^{+}\cdot\hat{\mathbf{p}}^{-}(t)+\left(\overline{\mathcal{G}}_{\mathrm{G}}^{-}\right)^{T}\cdot\hat{\mathbf{p}}^{+}(t)\right].
\end{aligned}
\tag{C10b}
$$

Next, we switch back to the Schrödinger picture, so that the master equation becomes:

$$
\partial_t\hat{\rho}_{\mathrm{S}} = -i\frac{1}{\hbar}\left[\hat{H}_{\mathrm{at}},\hat{\rho}_{\mathrm{S}}\right]+\mathcal{L}\hat{\rho}_{\mathrm{S}}, \tag{C11}
$$

where $\hat{H}_{\mathrm{at}}$ is the atom Hamiltonian, $\hat{\rho}_{\mathrm{S}} = e^{-i\frac{\hat{H}_{\mathrm{at}}}{\hbar}t}\hat{\rho}_{\mathrm{S,I}}e^{-i\frac{\hat{H}_{\mathrm{at}}}{\hbar}t}$ is the reduced density matrix in the Schrödinger picture. The Lindblad operator in the Schrödinger picture satisfies:

$$\mathcal{L}\hat{\rho}_{\mathrm{S}}=-\frac{i}{\hbar}\hat{\rho}_{\mathrm{S}}(t)\left[\hat{\mathbf{p}}^{-}\cdot\left(\overline{\boldsymbol{\mathcal{G}}}_{\mathrm{L}}^{-*}-\overline{\boldsymbol{\mathcal{G}}}_{\mathrm{G}}^{+T}\right)\cdot\hat{\mathbf{p}}+\hat{\mathbf{p}}^{+}\cdot\left(\overline{\boldsymbol{\mathcal{G}}}_{\mathrm{L}}^{+\dagger}-\overline{\boldsymbol{\mathcal{G}}}_{\mathrm{G}}^{-}\right)\cdot\hat{\mathbf{p}}\right]$$
$$+\frac{i}{\hbar}\left[\hat{\mathbf{p}}\cdot\left(\overline{\boldsymbol{\mathcal{G}}}_{\mathrm{L}}^{+}-\overline{\boldsymbol{\mathcal{G}}}_{\mathrm{G}}^{-\dagger}\right)\cdot\hat{\mathbf{p}}^{-}+\hat{\mathbf{p}}\cdot\left(\overline{\boldsymbol{\mathcal{G}}}_{\mathrm{L}}^{-T}-\overline{\boldsymbol{\mathcal{G}}}_{\mathrm{G}}^{+*}\right)\cdot\hat{\mathbf{p}}^{+}\right]\hat{\rho}_{\mathrm{S}}(t)$$
$$-\frac{i}{\hbar}\left[\hat{\mathbf{p}}^{-}\cdot\left(\overline{\boldsymbol{\mathcal{G}}}_{\mathrm{L}}^{+T}-\overline{\boldsymbol{\mathcal{G}}}_{\mathrm{G}}^{-*}\right)\hat{\rho}_{\mathrm{S}}(t)\cdot\hat{\mathbf{p}}+\hat{\mathbf{p}}^{+}\cdot\left(\overline{\boldsymbol{\mathcal{G}}}_{\mathrm{L}}^{-}-\overline{\boldsymbol{\mathcal{G}}}_{\mathrm{G}}^{+\dagger}\right)\hat{\rho}_{\mathrm{S}}(t)\cdot\hat{\mathbf{p}}\right] \tag{C12}$$
$$+\frac{i}{\hbar}\left[\hat{\mathbf{p}}\cdot\left(\overline{\boldsymbol{\mathcal{G}}}_{\mathrm{L}}^{-\dagger}-\overline{\boldsymbol{\mathcal{G}}}_{\mathrm{G}}^{+}\right)\hat{\rho}_{\mathrm{S}}(t)\cdot\hat{\mathbf{p}}^{-}+\hat{\mathbf{p}}\cdot\left(\overline{\boldsymbol{\mathcal{G}}}_{\mathrm{L}}^{+*}-\overline{\boldsymbol{\mathcal{G}}}_{\mathrm{G}}^{-T}\right)\hat{\rho}_{\mathrm{S}}(t)\cdot\hat{\mathbf{p}}^{+}\right]$$

Next, we note that the term $\frac{1}{\omega+\omega_{\mathrm{a}}^{*}}$ in the integrand of $\overline{\boldsymbol{\mathcal{G}}}_{\alpha}^{-}$ has no singularities in the integration range. This means that $\omega_{\mathrm{a}}$ can be taken as real-valued. In particular, it follows that $\overline{\boldsymbol{\mathcal{G}}}_{\alpha}^{-}$ is a Hermitian tensor:

$$\overline{\boldsymbol{\mathcal{G}}}_{\alpha}^{-}=\left(\overline{\boldsymbol{\mathcal{G}}}_{\alpha}^{-}\right)^{\dagger},\qquad \alpha=\mathrm{L,G}. \tag{C13}$$

Taking also into account that for a two-level system $\hat{\mathbf{p}}_{i}^{-}\hat{\mathbf{p}}_{j}^{-}=\hat{\mathbf{p}}_{i}^{+}\hat{\mathbf{p}}_{j}^{+}=0$, one finds that the Lindblad operator can be expressed as:

$$\mathcal{L}\hat{\rho}_{\mathrm{S}}=-\frac{i}{\hbar}\hat{\rho}_{\mathrm{S}}(t)\left[\hat{\mathbf{p}}^{-}\cdot\left(\overline{\boldsymbol{\mathcal{G}}}_{\mathrm{L}}^{-}-\overline{\boldsymbol{\mathcal{G}}}_{\mathrm{G}}^{+}\right)^{T}\cdot\hat{\mathbf{p}}^{+}+\hat{\mathbf{p}}^{+}\cdot\left(\overline{\boldsymbol{\mathcal{G}}}_{\mathrm{L}}^{+\dagger}-\overline{\boldsymbol{\mathcal{G}}}_{\mathrm{G}}^{-}\right)\cdot\hat{\mathbf{p}}^{-}\right]$$
$$+\frac{i}{\hbar}\left[\hat{\mathbf{p}}^{+}\cdot\left(\overline{\boldsymbol{\mathcal{G}}}_{\mathrm{L}}^{+}-\overline{\boldsymbol{\mathcal{G}}}_{\mathrm{G}}^{-}\right)\cdot\hat{\mathbf{p}}^{-}+\hat{\mathbf{p}}^{-}\cdot\left(\overline{\boldsymbol{\mathcal{G}}}_{\mathrm{L}}^{-}-\overline{\boldsymbol{\mathcal{G}}}_{\mathrm{G}}^{+\dagger}\right)^{T}\cdot\hat{\mathbf{p}}^{+}\right]\hat{\rho}_{\mathrm{S}}(t)$$
$$-\frac{i}{\hbar}\left[\hat{\mathbf{p}}^{-}\cdot\left(\overline{\boldsymbol{\mathcal{G}}}_{\mathrm{L}}^{+}-\overline{\boldsymbol{\mathcal{G}}}_{\mathrm{L}}^{+\dagger}\right)^{T}\hat{\rho}_{\mathrm{S}}(t)\cdot\hat{\mathbf{p}}^{+}+\hat{\mathbf{p}}^{+}\cdot\left(\overline{\boldsymbol{\mathcal{G}}}_{\mathrm{G}}^{+}-\overline{\boldsymbol{\mathcal{G}}}_{\mathrm{G}}^{+\dagger}\right)\hat{\rho}_{\mathrm{S}}(t)\cdot\hat{\mathbf{p}}^{-}\right]$$
$$+\frac{i}{\hbar}\left[\hat{\mathbf{p}}^{-}\cdot\left(\overline{\boldsymbol{\mathcal{G}}}_{\mathrm{L}}^{-}-\overline{\boldsymbol{\mathcal{G}}}_{\mathrm{L}}^{+T}+\overline{\boldsymbol{\mathcal{G}}}_{\mathrm{G}}^{-T}-\overline{\boldsymbol{\mathcal{G}}}_{\mathrm{G}}^{+}\right)\hat{\rho}_{\mathrm{S}}(t)\cdot\hat{\mathbf{p}}^{-}+\hat{\mathbf{p}}^{+}\cdot\left(\overline{\boldsymbol{\mathcal{G}}}_{\mathrm{L}}^{+*}-\overline{\boldsymbol{\mathcal{G}}}_{\mathrm{L}}^{-}+\overline{\boldsymbol{\mathcal{G}}}_{\mathrm{G}}^{+\dagger}-\overline{\boldsymbol{\mathcal{G}}}_{\mathrm{G}}^{-T}\right)\hat{\rho}_{\mathrm{S}}(t)\cdot\hat{\mathbf{p}}^{+}\right]$$

(C14)

It is useful to introduce the operator:

$$\hat{H}_{\mathrm{int,ef}}=-\hat{\mathbf{p}}^{+}\cdot\left[\frac{1}{2}\left(\overline{\boldsymbol{\mathcal{G}}}_{\mathrm{L}}^{+}+\overline{\boldsymbol{\mathcal{G}}}_{\mathrm{L}}^{+\dagger}\right)-\overline{\boldsymbol{\mathcal{G}}}_{\mathrm{G}}^{-}\right]\cdot\hat{\mathbf{p}}^{-}-\hat{\mathbf{p}}^{-}\cdot\left[\overline{\boldsymbol{\mathcal{G}}}_{\mathrm{L}}^{-}-\frac{1}{2}\left(\overline{\boldsymbol{\mathcal{G}}}_{\mathrm{G}}^{+}+\overline{\boldsymbol{\mathcal{G}}}_{\mathrm{G}}^{+\dagger}\right)\right]^{T}\cdot\hat{\mathbf{p}}^{+}. \tag{C15}$$

Then, the terms in the 1st and 2nd lines in the right-hand side of Eq. (C14) can be conveniently rewritten in terms of a commutator ($\left[\hat{A},\hat{B}\right]_{-}=\hat{A}\hat{B}-\hat{B}\hat{A}$) and an anti-commutator ($\left\{\hat{A},\hat{B}\right\}_{+}=\hat{A}\hat{B}+\hat{B}\hat{A}$), as follows:

$$\begin{aligned}\mathcal{L}\hat{\rho}_{\mathrm{S}}=&-i\frac{1}{\hbar}\left[\hat{H}_{\mathrm{int,ef}},\hat{\rho}_{\mathrm{S}}\right]_{-}+\frac{i}{2\hbar}\left\{\hat{\mathbf{p}}^{+}\cdot\left(\overline{\boldsymbol{\mathcal{G}}}_{\mathrm{L}}^{+}-\overline{\boldsymbol{\mathcal{G}}}_{\mathrm{L}}^{+\dagger}\right)\cdot\hat{\mathbf{p}}^{-}+\hat{\mathbf{p}}^{-}\cdot\left(\overline{\boldsymbol{\mathcal{G}}}_{\mathrm{G}}^{+}-\overline{\boldsymbol{\mathcal{G}}}_{\mathrm{G}}^{+\dagger}\right)^{T}\cdot\hat{\mathbf{p}}^{+},\hat{\rho}_{\mathrm{S}}(t)\right\}_{+}\\&-\frac{i}{\hbar}\left[\hat{\mathbf{p}}^{-}\cdot\left(\overline{\boldsymbol{\mathcal{G}}}_{\mathrm{L}}^{+}-\overline{\boldsymbol{\mathcal{G}}}_{\mathrm{L}}^{+\dagger}\right)^{T}\hat{\rho}_{\mathrm{S}}(t)\cdot\hat{\mathbf{p}}^{+}+\hat{\mathbf{p}}^{+}\cdot\left(\overline{\boldsymbol{\mathcal{G}}}_{\mathrm{G}}^{+}-\overline{\boldsymbol{\mathcal{G}}}_{\mathrm{G}}^{+\dagger}\right)\hat{\rho}_{\mathrm{S}}(t)\cdot\hat{\mathbf{p}}^{-}\right]\\&+\frac{i}{\hbar}\left[\hat{\mathbf{p}}^{-}\cdot\left(\overline{\boldsymbol{\mathcal{G}}}_{\mathrm{L}}^{-}-\overline{\boldsymbol{\mathcal{G}}}_{\mathrm{L}}^{+T}+\overline{\boldsymbol{\mathcal{G}}}_{\mathrm{G}}^{-T}-\overline{\boldsymbol{\mathcal{G}}}_{\mathrm{G}}^{+}\right)\hat{\rho}_{\mathrm{S}}(t)\cdot\hat{\mathbf{p}}^{-}+\hat{\mathbf{p}}^{+}\cdot\left(\overline{\boldsymbol{\mathcal{G}}}_{\mathrm{L}}^{+*}-\overline{\boldsymbol{\mathcal{G}}}_{\mathrm{L}}^{-}+\overline{\boldsymbol{\mathcal{G}}}_{\mathrm{G}}^{+\dagger}-\overline{\boldsymbol{\mathcal{G}}}_{\mathrm{G}}^{-T}\right)\hat{\rho}_{\mathrm{S}}(t)\cdot\hat{\mathbf{p}}^{+}\right].\end{aligned}\tag{C16}$$

In summary, the Lindblad operator for a two-level system interacting with an environment with gain is defined as shown above with the coefficients of the master equation determined by Eqs. (C8)-(C9).

*C. Secular approximation*

To proceed further, we apply the secular (rotating wave) approximation to Eq. (C16), retaining only the terms that preserve the resonance condition while discarding counter-rotating contributions in the Lindblad operator. This approximation leads to the standard Lindblad form, which guarantees the complete positivity and trace preservation of the density matrix evolution [46, 47]:

$$\begin{aligned}\mathcal{L}_{\mathrm{RW}}\hat{\rho}_{\mathrm{S}}=&-i\frac{1}{\hbar}\left[\hat{H}_{\mathrm{int,ef}},\hat{\rho}_{\mathrm{S}}\right]_{-}-\frac{1}{\hbar}\left\{\hat{\mathbf{p}}^{+}\cdot\overline{\boldsymbol{\mathcal{G}}}_{\mathrm{L}}^{\mathrm{NH}}\cdot\hat{\mathbf{p}}^{-}+\hat{\mathbf{p}}^{-}\cdot\left(\overline{\boldsymbol{\mathcal{G}}}_{\mathrm{G}}^{\mathrm{NH}}\right)^{T}\cdot\hat{\mathbf{p}}^{+},\hat{\rho}_{\mathrm{S}}(t)\right\}_{+}\\&+\frac{2}{\hbar}\hat{\mathbf{p}}^{-}\cdot\left(\overline{\boldsymbol{\mathcal{G}}}_{\mathrm{L}}^{\mathrm{NH}}\right)^{T}\hat{\rho}_{\mathrm{S}}(t)\cdot\hat{\mathbf{p}}^{+}+\frac{2}{\hbar}\hat{\mathbf{p}}^{+}\cdot\overline{\boldsymbol{\mathcal{G}}}_{\mathrm{G}}^{\mathrm{NH}}\hat{\rho}_{\mathrm{S}}(t)\cdot\hat{\mathbf{p}}^{-}.\end{aligned}\tag{C17}$$

Note that $\hat{H}_{\mathrm{int,ef}}$ is responsible for a Lamb shift, while the remaining terms, written in terms of the anti-Hermitian parts of the Green's functions,

$$\overline{\mathcal{G}}_{\alpha}^{\mathrm{NH}}=\frac{\overline{\mathcal{G}}_{\alpha}^{+}-\overline{\mathcal{G}}_{\alpha}^{+\dagger}}{2i}, \qquad \alpha=\mathrm{L,G}. \tag{C18}$$

are responsible for the irreversible-type (e.g., spontaneous decay) dynamics.

To confirm that Eq. (C17) is indeed compatible with the Lindblad form, we need to show that $\overline{\mathcal{G}}_{\alpha}^{\mathrm{NH}}$ is non-negative. To this end, we combine Eq. (C9) with the Sokhotski–Plemelj identity ($\frac{1}{\omega-0^{+}i}=\mathrm{PV}\frac{1}{\omega}+i\pi\delta(\omega)$) to obtain:

$$\overline{\mathcal{G}}_{\alpha}^{\mathrm{NH}}=\frac{1}{2i}\mathbf{\Lambda}_{\alpha}\left(\omega_{\mathrm{a}}\right), \quad \alpha=\mathrm{L,G}. \tag{C19}$$

We used the identity $\mathbf{\Lambda}_{\alpha}^{\dagger}=-\mathbf{\Lambda}_{\alpha}$. Substituting now Eq. (C8) into the above result, we obtain the following explicit formula for the coefficients of the quantum master equation:

$$\overline{\mathcal{G}}_{\mathrm{L/G}}^{\mathrm{NH}}=\pm\int d^{3}\mathbf{r}_{0}\,\overline{\mathcal{G}}\left(\mathbf{r}_{\mathrm{a}},\mathbf{r}_{0},\omega_{\mathrm{a}}\right)\cdot\mathbf{M}_{\mathrm{L/G}}''\left(\mathbf{r}_{0}\right)\cdot\overline{\mathcal{G}}^{\dagger}\left(\mathbf{r}_{\mathrm{a}},\mathbf{r}_{0},\omega_{\mathrm{a}}\right). \tag{C20}$$

The right-hand side is evidently a non-negative matrix because $\pm\mathbf{M}_{\mathrm{L/G}}''$ are non-negative matrices. This confirms that $\mathcal{L}_{\mathrm{RW}}$ has indeed the Lindblad form, as we wanted to show.

It is useful to note that from Eqs. (A7), (C20) and $\mathbf{M}''=\mathbf{M}_{\mathrm{L}}''+\mathbf{M}_{\mathrm{G}}''$, we obtain the useful identity:

$$\overline{\mathcal{G}}_{\mathrm{L}}^{\mathrm{NH}}-\overline{\mathcal{G}}_{\mathrm{G}}^{\mathrm{NH}}=\frac{\overline{\mathcal{G}}-\overline{\mathcal{G}}^{\dagger}}{2i}, \tag{C21}$$

with $\overline{\mathcal{G}}$ evaluated at $\mathbf{r}=\mathbf{r}'=\mathbf{r}_{\mathrm{a}}$ and $\omega=\omega_{\mathrm{a}}$. The above identity also confirms that for a passive system ($\overline{\mathcal{G}}_{\mathrm{G}}^{\mathrm{NH}}=0$) the Lindblad operator reduces to the result reported in previous works [50-51].

## Appendix D: Temperature effect

In this appendix, we extend the Lindblad master equation to scenarios where the environment is characterized by a generalized Gibbs state ($\hat{\rho}_{\mathrm{E}} \approx \hat{\rho}_{\mathrm{E,Gibbs}}$). The key difference from the analysis in Appendix C is that the field correlations $\mathrm{tr}\left\{\hat{\rho}_{\mathrm{E}}\hat{\mathbf{F}}\left(\mathbf{r},t\right)\otimes\hat{\mathbf{F}}\left(\mathbf{r}',t-\tau\right)\right\}$ are no longer determined by Eq. (B3) but instead by Eq. (B9).

The final result in Eq. (B9) is the sum of the four terms presented in Eq. (B8). The terms in Eqs. (B8a) and (B8d) correspond to the first and second terms in Eq. (B3), respectively, differing only by the additional coefficient $1+N_{\omega}$. Consequently, these two terms originate contributions to the Lindblad operator with the same structure as those derived in Appendix C.

Specifically, let us focus on the non-conservative terms of the Lindblad operator, i.e., the part that does not depend on the term $\hat{H}_{\mathrm{int,ef}}$. As shown in Eq. (C20), the coefficients of the non-conservative terms are all evaluated at the atomic transition frequency $\omega_{\mathrm{a}}$. Therefore, we conclude that the contributions of the terms in Eqs. (B8a) and (B8d) to the non-conservative part of the Lindblad operator remain the same as in the vacuum case (Appendix C), apart from a multiplicative coefficient identical to $1+N_{\omega_{\mathrm{a}}}$.

Next, we analyze the contributions of the terms in Eqs. (B8b) and (B8c) to the Lindblad operator. By comparing Eqs. (B8b) and (B8d), we observe that they share the same structure, except that the indices L and G are interchanged and the leading coefficient $1+N_{\omega_{\mathrm{a}}}$ is replaced by $-N_{\omega_{\mathrm{a}}}$. The same observation applies to the terms in Eqs. (B8a) and (B8c). Consequently, the contributions of the terms in Eqs. (B8b) and (B8c) to the Lindblad operator are structurally similar to those of the terms in Eqs. (B8a) and (B8d), but with the indices L and G interchanged and the term $1+N_{\omega_\mathrm{a}}$ is replaced by $N_{\omega_\mathrm{a}}$. The minus sign is accounted for by the difference in the definition of the coefficients in Eq. (B2); in other words, interchanging the indices L and G naturally absorbs the extra minus sign.

The preceding analysis demonstrates that the non-conservative part of the Lindblad operator for an environment described by a generalized Gibbs state (thermal bath) is given by (compare with Eq. (C17)):

$$\mathcal{L}_\mathrm{d}\hat{\rho}_\mathrm{S} = -\frac{1}{\hbar}\left\{\hat{\mathbf{p}}^+ \cdot \left[\left(1+N_{\omega_\mathrm{a}}\right)\overline{\boldsymbol{\mathcal{G}}}_\mathrm{L}^\mathrm{NH} + N_{\omega_\mathrm{a}}\overline{\boldsymbol{\mathcal{G}}}_\mathrm{G}^\mathrm{NH}\right]\cdot\hat{\mathbf{p}}^- + \hat{\mathbf{p}}^- \cdot \left[\left(1+N_{\omega_\mathrm{a}}\right)\overline{\boldsymbol{\mathcal{G}}}_\mathrm{G}^\mathrm{NH} + N_{\omega_\mathrm{a}}\overline{\boldsymbol{\mathcal{G}}}_\mathrm{L}^\mathrm{NH}\right]^T \cdot \hat{\mathbf{p}}^+, \hat{\rho}_\mathrm{S}(t)\right\}_+ + \frac{2}{\hbar}\hat{\mathbf{p}}^- \cdot \left[\left(1+N_{\omega_\mathrm{a}}\right)\overline{\boldsymbol{\mathcal{G}}}_\mathrm{L}^\mathrm{NH} + N_{\omega_\mathrm{a}}\overline{\boldsymbol{\mathcal{G}}}_\mathrm{G}^\mathrm{NH}\right]^T \hat{\rho}_\mathrm{S}(t)\cdot\hat{\mathbf{p}}^+ + \frac{2}{\hbar}\hat{\mathbf{p}}^+ \cdot \left[\left(1+N_{\omega_\mathrm{a}}\right)\overline{\boldsymbol{\mathcal{G}}}_\mathrm{G}^\mathrm{NH} + N_{\omega_\mathrm{a}}\overline{\boldsymbol{\mathcal{G}}}_\mathrm{L}^\mathrm{NH}\right]\hat{\rho}_\mathrm{S}(t)\cdot\hat{\mathbf{p}}^-. \tag{D1}$$

From the discussion in Appendix C, it is evident that all the matrices in the above equation (e.g., $\left(1+N_{\omega_\mathrm{a}}\right)\overline{\boldsymbol{\mathcal{G}}}_\mathrm{L}^\mathrm{NH}$) are nonnegative and Hermitian. This property ensures that the terms associated with these matrices retain the Lindblad form [46]. Furthermore, our result regarding the temperature's effect on the Lindblad dissipator aligns with previous studies [51, 52].

The conservative part of the Lindblad operator $\hat{H}_\mathrm{int,ef}$ in Eq. (C15) can be extended to the thermal case using similar considerations. However, in this case, the temperature effect is not solely determined by $N_{\omega_\mathrm{a}}$ but also requires modifications to the definitions of the Green's functions. This aspect will not be explored in this work.

## Appendix E: Green's function

In this Appendix, we calculate the tensors $\overline{\boldsymbol{\mathcal{G}}}_{\alpha}^{\mathrm{NH}}$ using a quasi-static approximation, i.e., neglecting retardation effects.

*A. Quasi-static Approximation*

The system's Green's function is a 6×6 tensor that satisfies Eq. (A2). For systems with a trivial magnetic response, it is sufficient to consider only the electric part of the Green's function, which is a 3×3 tensor that satisfies:

$$\nabla\times\nabla\times\overline{\boldsymbol{\mathcal{G}}}_{\mathrm{e}}-\left(\frac{\omega}{c}\right)^{2}\overline{\varepsilon}\cdot\overline{\boldsymbol{\mathcal{G}}}_{\mathrm{e}}=\frac{1}{\varepsilon_{0}}\left(\frac{\omega}{c}\right)^{2}\mathbf{1}_{3\times3}\delta\left(\mathbf{r}-\mathbf{r}'\right) \tag{E1}$$

with $\overline{\varepsilon}$ the relative permittivity tensor.

In the quasi-static regime, we seek solutions such that $\overline{\boldsymbol{\mathcal{G}}}_{\mathrm{e}}\cdot\hat{\mathbf{u}}_{i}=-\nabla\phi_{i}$ ($i$=1,2,3), with $\phi_i$ a (normalized) electric potential due to a point dipole. Substituting this formula into Eq. (E1) and taking the divergence of both sides, we find that the electric potential satisfies $\nabla\cdot\left(\overline{\varepsilon}\cdot\nabla\phi_{i}\right)=\frac{1}{\varepsilon_{0}}\frac{\partial}{\partial x_{i}}\delta\left(\mathbf{r}-\mathbf{r}'\right)=-\frac{1}{\varepsilon_{0}}\frac{\partial}{\partial x_{i}'}\delta\left(\mathbf{r}-\mathbf{r}'\right)$. Hence, $\phi_{i}=\frac{\partial\Phi}{\partial x_{i}'}$, with $\Phi$ the electric potential due to a point charge:

$$\nabla\cdot\left(\overline{\varepsilon}\cdot\nabla\Phi\right)=-\frac{1}{\varepsilon_{0}}\delta\left(\mathbf{r}-\mathbf{r}'\right). \tag{E2}$$

The Green's function can be expressed as:

$$\overline{\boldsymbol{\mathcal{G}}}_{\mathrm{e}}=-\overrightarrow{\nabla}\Phi\overleftarrow{\nabla}', \tag{E3}$$

where the arrow above the nabla operator indicates the direction on which the spatial derivatives act.

For simplicity, in this article we restrict our attention to isotropic systems, where the permittivity is a scalar quantity. Furthermore, we assume that the materials are invariant under translations along the $x$ and $y$ directions. The region $z>0$ is filled with air, while the region $z<0$ is occupied by a medium of interest with permittivity $\varepsilon$. Then, using Fourier methods, it can be readily shown that when the source point $\mathbf{r}'$ is located in the material ( $z'<0$ ), the electric potential can be written as:

$$\Phi = \frac{1}{(2\pi)^2}\iint d^2\mathbf{k}_{\parallel} e^{i\mathbf{k}_{\parallel}\cdot(\mathbf{r}-\mathbf{r}')} g(z,z'), \qquad \text{with} \tag{E4a}$$

$$g(z,z') = \frac{1}{\varepsilon_0 \varepsilon}\frac{1}{2k_{\parallel}}\begin{cases} e^{-k_{\parallel}|z-z'|} + Re^{k_{\parallel}z'}e^{k_{\parallel}z}, & z<0 \\ e^{k_{\parallel}z'}Te^{-k_{\parallel}z}, & z>0 \end{cases}. \tag{E4b}$$

Here, $\mathbf{k}_{\parallel} = (k_x, k_y, 0)$ and $R$ and $T$ are reflection and transmission coefficients which can be found by imposing the continuity of the electric potential and displacement vector at the interface:

$$R = \frac{\varepsilon - 1}{\varepsilon + 1}, \qquad T = 1 + R = \frac{2\varepsilon}{\varepsilon + 1}. \tag{E5}$$

In particular, when the observation point is on the air side ( $z>0$ and $z'<0$ ) the Green's function can be expressed as:

$$\overline{\mathcal{G}}_{\mathrm{e}} = \frac{1}{(2\pi)^2 \varepsilon_0}\iint d^2\mathbf{k}_{\parallel} \mathcal{G}_{\mathbf{k}_{\parallel}} e^{i\mathbf{k}_{\parallel}\cdot(\mathbf{r}-\mathbf{r}')}, \qquad \text{with} \tag{E6a}$$

$$\overline{\mathcal{G}}_{\mathbf{k}_{\parallel}} = \left(i\mathbf{k}_{\parallel} - k_{\parallel}\hat{\mathbf{z}}\right) \otimes \left(i\mathbf{k}_{\parallel} - k_{\parallel}\hat{\mathbf{z}}\right) e^{k_{\parallel}z'} e^{-k_{\parallel}z} \frac{1}{k_{\parallel}}\frac{1}{\varepsilon + 1}. \tag{E6b}$$

*B. Calculation of $\overline{\mathcal{G}}_{\alpha}^{\mathrm{NH}}$ for an isotropic gain medium*

Next, we obtain explicit formulas for $\overline{\boldsymbol{\mathcal{G}}}_{\alpha}^{\mathrm{NH}}$ by evaluating the integrals in Eq. (14). Taking into account that $\mathbf{M}''$ can be identified with a scalar ($\varepsilon_0\varepsilon''$), it can be readily shown with the help of Eq. (E6) that the electric-part of $\overline{\boldsymbol{\mathcal{G}}}_{\alpha}^{\mathrm{NH}}$ is:

$$\overline{\boldsymbol{\mathcal{G}}}_{\mathrm{L,G}}^{\mathrm{NH}} = \pm\frac{1}{(2\pi)^2\varepsilon_0}\int_{-\infty}^{0} dz' \iint d^2\mathbf{k}_{\parallel}\, \varepsilon''_{\mathrm{L,G}}(\omega_{\mathrm{a}})\, \boldsymbol{\mathcal{G}}_{\mathbf{k}_{\parallel}} \cdot \left(\boldsymbol{\mathcal{G}}_{\mathbf{k}_{\parallel}}\right)^{\dagger}. \tag{E7}$$

We used the decomposition of the loss function $\varepsilon'' = \varepsilon''_{\mathrm{L}} + \varepsilon''_{\mathrm{G}} = \varepsilon''_{\mathrm{L}} - \left|\varepsilon''_{\mathrm{G}}\right|$ in dissipative and gain components. The integration over $z'$ can be readily carried out to obtain:

$$\overline{\boldsymbol{\mathcal{G}}}_{\mathrm{L,G}}^{\mathrm{NH}} = \frac{-1}{(2\pi)^2\varepsilon_0}\iint d^2\mathbf{k}_{\parallel} \frac{2\left|\varepsilon''_{\mathrm{L,G}}\right|}{\left|\varepsilon+1\right|^2}\frac{e^{-2k_{\parallel}z_{\mathrm{a}}}}{2k_{\parallel}}\left(i\mathbf{k}_{\parallel} - k_{\parallel}\hat{\mathbf{z}}\right)\otimes\left(i\mathbf{k}_{\parallel} + k_{\parallel}\hat{\mathbf{z}}\right). \tag{E8}$$

It is implicit that the permittivity is evaluated at the atomic transition frequency $\omega_{\mathrm{a}}$. In the above, $z_{\mathrm{a}}$ stands for the atom position above the dielectric substrate. By comparing Eq. (E8) with the well-known spectral representation of the function $1/\left(4\pi\left|\mathbf{r}\right|\right)$, one readily sees that $\overline{\boldsymbol{\mathcal{G}}}_{\mathrm{L,G}}^{\mathrm{NH}}$ can be expressed as:

$$\overline{\boldsymbol{\mathcal{G}}}_{\mathrm{L,G}}^{\mathrm{NH}} = \frac{2\left|\varepsilon''_{\mathrm{L,G}}\right|}{\left|\varepsilon+1\right|^2}\frac{1}{\varepsilon_0}\vec{\nabla}\frac{1}{4\pi\sqrt{(x-x')^2+(y-y')^2+(z+z')^2}}\overleftarrow{\nabla}', \tag{E9}$$

where the right-hand side is evaluated with $z = z' = z_{\mathrm{a}}$, $x = x'$ and $y = y'$. Hence, we conclude that:

$$\overline{\boldsymbol{\mathcal{G}}}_{\mathrm{L,G}}^{\mathrm{NH}} = \frac{\left|\varepsilon''_{\mathrm{L,G}}\right|}{\left|\varepsilon+1\right|^2}\frac{1}{16\pi\varepsilon_0 z_{\mathrm{a}}^3}\left(\mathbf{1}_t + 2\hat{\mathbf{z}}\otimes\hat{\mathbf{z}}\right), \tag{E10}$$

with $\mathbf{1}_t = \hat{\mathbf{x}}\otimes\hat{\mathbf{x}} + \hat{\mathbf{y}}\otimes\hat{\mathbf{y}}$.

For consistency, we note that, when both the source and observation points on the air region, a similar analysis shows that the quasi-static system Green's function can be written as:

$$\overline{\mathcal{G}}_{\mathrm{e}} = -\frac{1}{\varepsilon_0}\vec{\nabla}\left[\frac{1}{4\pi\sqrt{\left(x-x'\right)^2+\left(y-y'\right)^2+\left(z-z'\right)^2}}+\frac{\tilde{R}}{4\pi\sqrt{\left(x-x'\right)^2+\left(y-y'\right)^2+\left(z+z'\right)^2}}\right]\overleftarrow{\nabla}'$$

(E11)

Here, $\tilde{R} = \frac{1-\varepsilon}{1+\varepsilon}$ is the quasi-static reflection coefficient for illumination from the air side. Taking both the source and observation points coincident with the atom location ($\mathbf{r} = \mathbf{r}_{\mathrm{a}} = \mathbf{r}'$), we find that

$$\frac{\overline{\mathcal{G}}_{\mathrm{e}} - \overline{\mathcal{G}}_{\mathrm{e}}^{\dagger}}{2i} = -\mathrm{Im}\left\{\tilde{R}\right\}\frac{1}{\varepsilon_0}\frac{\mathbf{1}_t + 2\hat{\mathbf{z}}\otimes\hat{\mathbf{z}}}{4\pi\left(2z_{\mathrm{a}}\right)^3} = \overline{\mathcal{G}}_{\mathrm{L}}^{\mathrm{NH}} - \overline{\mathcal{G}}_{\mathrm{G}}^{\mathrm{NH}}. \quad \text{(E12)}$$

In the rightmost identity we used $\varepsilon_{\mathrm{L}}'' - \left|\varepsilon_{\mathrm{G}}''\right| = \varepsilon_{\mathrm{L}}'' + \varepsilon_{\mathrm{G}}'' = \varepsilon''$ and $\mathrm{Im}\left\{\tilde{R}\right\} = -2\varepsilon''/\left|1+\varepsilon\right|^2$. The above formula agrees with the identity (C21), thereby confirming the internal consistency of the theory.

### *C. Calculation of $\overline{\mathcal{G}}_{\alpha}^{\mathrm{NH}}$ for a moving plasmonic slab*

Next, we evaluate $\overline{\mathcal{G}}_{\alpha}^{\mathrm{NH}}$ for the case of a moving plasmonic slab. In the lab frame, the slab moves with velocity $v$ along the *x*-direction with $v > 0$.For non-relativistic velocities the moving material response in the lab frame is described by the permittivity $\varepsilon\left(\omega'\right)$, with $\omega' = \omega - k_x v$ the frequency evaluated in the slab co-moving frame [20, 21]. As a result, the material response becomes spatially dispersive. As further discussed in Ref. [21], in this scenario the separation between the dissipative ($\mathrm{Im}\left\{\varepsilon\left(\omega'\right)\right\} > 0$ corresponding to $k_x < \omega / v$) and gain ($\operatorname{Im}\{\varepsilon(\omega')\} < 0$ corresponding to $k_x > \omega / v$) parts of the material response is carried out in the spectral domain. From these observations, it is straightforward to check that the analysis of the previous subsection, and in particular Eq. (E8), can be generalized to the case of the moving plasmonic slab as follows:

$$\overline{\mathcal{G}}_{\mathrm{L}}^{\mathrm{NH}} = \frac{1}{(2\pi)^2 \varepsilon_0} \int_{-\infty}^{\omega_{\mathrm{a}}/v} dk_x \int_{-\infty}^{+\infty} dk_y \, e^{-2k_{\|} z_{\mathrm{a}}} \left| \operatorname{Im}\{\tilde{R}\} \right| \frac{1}{2k_{\|}} \left( i\mathbf{k}_{\|} - k_{\|}\hat{\mathbf{z}} \right) \otimes \left( -i\mathbf{k}_{\|} - k_{\|}\hat{\mathbf{z}} \right). \quad \text{(E13a)}$$

$$\overline{\mathcal{G}}_{\mathrm{G}}^{\mathrm{NH}} = \frac{1}{(2\pi)^2 \varepsilon_0} \int_{\omega_{\mathrm{a}}/v}^{+\infty} dk_x \int_{-\infty}^{+\infty} dk_y \, e^{-2k_{\|} z_{\mathrm{a}}} \left| \operatorname{Im}\{\tilde{R}\} \right| \frac{1}{2k_{\|}} \left( i\mathbf{k}_{\|} - k_{\|}\hat{\mathbf{z}} \right) \otimes \left( -i\mathbf{k}_{\|} - k_{\|}\hat{\mathbf{z}} \right). \quad \text{(E13b)}$$

In the above we used $\operatorname{Im}\{\tilde{R}\} = -2\varepsilon'' / |1+\varepsilon|^2$ with $\tilde{R} = \frac{1-\varepsilon}{1+\varepsilon}$, being implicit that $\varepsilon$ stands for the Doppler shifted permittivity $\varepsilon(\omega_{\mathrm{a}} - k_x v)$.

We model the slab permittivity as a standard Drude model with infinitesimal dissipation, $\varepsilon(\omega) = 1 - 2\omega_{\mathrm{sp}}^2 / \left(\omega\left(\omega + 0^+ i\right)\right)$, where $\omega_{\mathrm{sp}}$ is the surface plasmon resonance. A straightforward analysis shows that $\left|\operatorname{Im}\{\tilde{R}(\omega)\}\right| = \frac{\pi\omega_{\mathrm{sp}}}{2}\delta(\omega - \omega_{\mathrm{sp}}) + \frac{\pi\omega_{\mathrm{sp}}}{2}\delta(\omega + \omega_{\mathrm{sp}})$.

Using this result, the integration in Eq. (E13) can be carried out analytically (it is implicit that *v* is positive):

$$\overline{\mathcal{G}}_{\mathrm{L,G}}^{\mathrm{NH}} = \frac{1}{(2\pi)^2 \varepsilon_0} \frac{\pi\omega_{\mathrm{sp}}}{4v} \int_{-\infty}^{+\infty} dk_y \, e^{-2k_{\|} z_{\mathrm{a}}} \frac{1}{k_{\|}} \left( i\mathbf{k}_{\|} - k_{\|}\hat{\mathbf{z}} \right) \otimes \left( -i\mathbf{k}_{\|} - k_{\|}\hat{\mathbf{z}} \right) \Bigg|_{k_x = \frac{\omega_{\mathrm{a}} \pm \omega_{\mathrm{sp}}}{v}}. \quad \text{(E14)}$$

The remaining integrals are analogous to those presented in Ref. [68] and can be analytically evaluated in terms of modified Bessel functions of the second kind ($K_n$):

$$\overline{\mathcal{G}}_{\alpha}^{\mathrm{NH}}=\frac{1}{16\pi\varepsilon_0}k_{\alpha}^2\frac{\omega_{\mathrm{sp}}}{v}\begin{pmatrix}2K_0 & 0 & -i2\,\mathrm{sgn}\left(k_{\alpha}\right)K_1\\ 0 & K_2-K_0 & 0\\ +i2\,\mathrm{sgn}\left(k_{\alpha}\right)K_1 & 0 & K_2+K_0\end{pmatrix}. \quad \text{(E15)}$$

with $\alpha=\mathrm{L,G}$ and $k_{\mathrm{L}}=\frac{\omega_{\mathrm{a}}-\omega_{\mathrm{sp}}}{v}$ and $k_{\mathrm{G}}=\frac{\omega_{\mathrm{a}}+\omega_{\mathrm{sp}}}{v}$. The argument of the Bessel functions is $2\left|k_{\alpha}\right|z_{\mathrm{a}}$ and is omitted for conciseness.

For non-relativistic velocities and an atom placed at a subwavelength distance from the moving slab, we can safely assume that $2\left|k_{\alpha}\right|z_{\mathrm{a}}>>1$ when $\omega_{\mathrm{a}},\omega_{\mathrm{sp}}$ are sufficiently different. In such a case, we can use the large argument expansion $K_n\left(x\right)\approx\sqrt{\frac{\pi}{2x}}e^{-x}$ of the modified Bessel functions to obtain:

$$\overline{\mathcal{G}}_{\alpha}^{\mathrm{NH}}\approx\frac{1}{16\pi\varepsilon_0}k_{\alpha}^2\frac{\omega_{\mathrm{sp}}}{v}\sqrt{\frac{\pi}{\left|k_{\alpha}\right|z_{\mathrm{a}}}}e^{-2\left|k_{\alpha}\right|z_{\mathrm{a}}}\left(\hat{\mathbf{x}}+is_{\alpha}\hat{\mathbf{z}}\right)\otimes\left(\hat{\mathbf{x}}-is_{\alpha}\hat{\mathbf{z}}\right), \quad \text{(E16)}$$

with $s_{\alpha}=\mathrm{sgn}\,k_{\alpha}$ ($\alpha=\mathrm{L,G}$). Note that $s_{\mathrm{G}}=+1$ and $s_{\mathrm{L}}$ depends in the sign of $\omega_{\mathrm{a}}-\omega_{\mathrm{sp}}$.

## References

[1] K. Y. Bliokh, *et al*, “Roadmap on Structured Waves”, *J. Optics*, **25**, 103001, (1-77), (2023).

[2] I. V. Lindell, A. H. Sihvola, S. A. Tretyakov, and A. J. Viitanen, *Electromagnetic Waves in Chiral and Bi-Isotropic Media* (Artech House, Boston, MA, 1994).

[3] D. Vanderbilt, *Berry Phases in Electronic Structure Theory:Electric Polarization, Orbital Magnetization and Topological Insulators* (Cambridge University Press, Cambridge, UK, 2018).

[4] P. Lodahl, S. Mahmoodian, S. Stobbe, A. Rauschenbeutel, P. Schneeweiss, J. Volz, H. Pichler, P. Zoller, “Chiral quantum optics”, *Nature*, **541**, 473, (2017).

[5] D. G. Suárez-Forero, M. J. Mehrabad, C. Vega, A. González-Tudela, and M. Hafezi, “Chiral Quantum Optics: Recent Developments and Future Directions”, *PRX Quantum* **6**, 020101 (2025).

[6] I. Shomroni, S. Rosenblum, Y. Lovsky, O. Bechler, G. Guendelman, and B. Dayan, “All–optical routing of single photons by a one-atom switch controlled by a single photon”, *Science*, **345**, 903 (2014).

[7] C. Sayrin, C. Junge, R. Mitsch, B. Albrecht, D. O'Shea, P. Schneeweiss, J. Volz, and A. Rauschenbeutel, "Nanophotonic Optical Isolator Controlled by the Internal State of Cold Atoms", *Phys. Rev. X* **5**, 041036 (2015).

[8] M. Scheucher, A. Hilico, E. Will, J. Volz, A. Rauschenbeutel, "Quantum optical circulator controlled by a single chirally coupled atom", *Science*, **354**, 1577–1580 (2016).

[9] F. Tay, S. Sanders, A. Baydin, Z. Song, D. M. Welakuh, A. Alabastri, V. Rokaj, C. B. Dag, and J. Kono, "Terahertz chiral photonic-crystal cavities for Dirac gap engineering in graphene", *Nat. Commun.* **16**, 5270 (2025).

[10] R. R. Riso, M. Castagnola, E. Ronca, and H. Koch, Chiral polaritonics: cavity-mediated enantioselective excitation condensation, *Rep. Prog. Phys.* **88**, 027901 (2025).

[11] L. Feng, R. El-Ganainy, and L. Ge, Non-Hermitian photonics based on parity–time symmetry, *Nat. Photonics* **11**, 752 (2017).

[12] M. A. Miri and A. Alù, "Exceptional points in optics and photonics", *Science* **363**, eaar7709 (2019).

[13] E. Galiffi, R. Tirole, S. Yin, H. Li, S. Vezzoli, P. A. Huidobro, M. G. Silveirinha, R. Sapienza, A. Alù, J. B. Pendry, "Photonics of Time-Varying Media", *Adv. Photon.* **4**, 014002, (2022), doi: 10.1117/1.AP.4.1.014002

[14] K. Ding, C. Fang, G. Ma, "Non-Hermitian topology and exceptional-point geometries," *Nat. Rev. Phys.* **4**, 745 (2022).

[15] S. Lannebère, D. E. Fernandes, T. A. Morgado, M. G. Silveirinha, "Nonreciprocal and non-Hermitian material response inspired by semiconductor transistors", *Phys. Rev. Lett.*, **128**, 013902, (2022).

[16] S. Buddhiraju, A. Song, G. T. Papadakis, and S. Fan, "Nonreciprocal metamaterial obeying time-reversal symmetry", *Phys. Rev. Lett.* **124**, 257403 (2020).

[17] T. G. Rappoport, T. A. Morgado, S. Lannebère, M. G. Silveirinha, "Engineering transistor-like optical gain in two-dimensional materials with Berry curvature dipoles", *Phys. Rev. Lett.*, **130**, 076901, (2023).

[18] T. A. Morgado, T. G. Rappoport, S. S. Tsirkin, S. Lannebère, I. Souza, M. G. Silveirinha, "Non-Hermitian Linear Electrooptic Effect in 3D materials", *Phys. Rev. B*, **109**, 245126, 2024.

[19] S. Lannebère, T. G. Rappoport, T. A. Morgado, I. Souza, and M. G. Silveirinha, "Symmetry Analysis of the Non-Hermitian Electro-Optic Effect in Crystals", arXiv:2502.03399, (2025).

[20] M. G. Silveirinha, "Optical instabilities and spontaneous light emission by polarizable moving matter", *Phys. Rev. X*, **4**, 031013, (2014).

[21] D. Oue, J. B. Pendry, M. G. Silveirinha, "Stable-to-unstable transition in quantum friction", *Phys. Rev. Res.*, **6**, 043074, (2024).

[22] R. Matloob, R. Loudon, M. Artoni, S. M. Barnett, and J. Jeffers, "Electromagnetic field quantization in amplifying dielectrics" *Phys. Rev. A* **55**, 1623 (1997).

[23] S. Scheel, L. Knöll, and D.-G. Welsch, "QED commutation relations for inhomogeneous Kramers-Kronig dielectrics", *Phys. Rev. A* **58**, 700 (1998).

[24] C. Raabe and D.-G. Welsch, "QED in arbitrary linear media: Amplifying media", *Eur. Phys. J. Special Topics*, **160**, 371 (2008).

[25] T. von Foerster and R. J. Glauber, "Quantum Theory of Light Propagation in Amplifying Media", *Phys. Rev. A* **3**, 1484 (1971).

[26] J. Jeffers, N. Imoto, and R. Loudon, "Quantum optics of traveling-wave attenuators and amplifiers", *Phys. Rev. A*, **47**, 3346 (1993).

[27] J. Jeffers, S. M. Barnett, R. Loudon, R. Matloob, and M. Artoni, "Canonical quantum theory of light propagation in amplifying media", *Optics Comm.,* **131**, 66 (1996).

[28] M. G. Silveirinha, "Quantization of the Electromagnetic Field in Non-dispersive Polarizable Moving Media above the Cherenkov Threshold", *Phys. Rev. A*, **88**, 043846, (2013).

[29] S. Lannebère, D. E. Fernandes, T. A. Morgado, M. G. Silveirinha, "Chiral-Gain Photonics", *Laser Photonics Rev.* 2400881 (1-17), (2025).

[30] E. Amooghorban, N. A. Mortensen, and M. Wubs "Quantum Optical Effective-Medium Theory for Loss-Compensated Metamaterials," *Phys. Rev. Lett.* **110** 153602 (2013).

[31] W. Vogel and D.-G. Welsch, Quantum Optics, 3rd ed.(John Wiley & Sons, 2006).

[32] S. Y. Buhmann, D. T. Butcher, and S. Scheel, "Macroscopic quantum electrodynamics in nonlocal and nonreciprocal media", *New J. Phys.* **14**, 083034 (2012).

[33] T. Holstein and H. Primakoff, Field Dependence of the Intrinsic Domain Magnetization of a Ferromagnet, *Phys. Rev.* **58**, 1098 (1940).

[34] M. G. Silveirinha, "Theory of Quantum Friction", *New J. Phys.*, **16**, 063011, (2014).

[35] M. Rigol, A. Muramatsu and M. Olshanii "Hard-core bosons on optical superlattices: dynamics, relaxation in the superfluid, insulating regimes", *Phys. Rev. A* **74** 053616 (2006).

[36] M. Rigol, V. Dunjko, V. Yurovsky, and M. Olshanii, "Relaxation in a completely integrable many-body quantum system: An ab initio study of the dynamics of the highly excited states of 1D lattice hard-core bosons", *Phys. Rev. Lett.* **98**, 050405 (2007).

[37] F. H. L. Essler, G. Mussardo, and M. Panfil "Generalized Gibbs ensembles for quantum field theories", *Phys. Rev. A* **91**, 051602(R) (2015).

[38] L. D'Alessio, Y. Kafri, A. Polkovnikov, and M. Rigol "From quantum chaos and eigenstate thermalization to statistical mechanics and thermodynamics", *Adv. Phys.* **65**, 239 (2016).

[39] G. V. Dedkov and A. A. Kyasov, "Fluctuation-electromagnetic interaction under dynamic and thermal nonequilibrium condition", *Phys. Usp.* **60**, 559 (2017).

[40] G. Tang, L. Zhang, Y. Zhang, J. Chen, and C. T. Chan, "Near-field energy transfer between graphene and magneto-optic media", *Phys. Rev. Lett.* **127**, 247401 (2021).

[41] L. D. Landau and E. M. Lifshitz, Statistical Physics: Part 2 (Pergamon Press, Oxford, 1981).

[42] S. M. Rytov, Yu A. Kravtsov and V. I. Tatarskii, Principles of Statistical Radiophysics, Vol. 3, (Springer, Berlin, 1989).

[43] D. Oue, B. Shapiro, M. G. Silveirinha, "Quantum Friction near the Instability Threshold", *Phys. Rev. B*, **111**, 075403, (2025).

[44] J. B. Pendry, "Shearing the vacuum-quantum friction", *J. Phys. Condens. Matter* **9**, 10301 (1997).

[45] A. Volokitin and B. Persson, "Theory of friction: the contribution from a fluctuating electromagnetic field", *J. Phys. Condens. Matter* **11**, 345 (1999).

[46] G. Lindblad, "On the generators of quantum dynamical semigroups", *Commun.Math. Phys.* **48**, 119 (1976).

[47] H.-P. Breuer and F. Petruccione, *The Theory of Open Quantum Systems*, Oxford University Press (2002).

[48] F. Campaioli, J. H. Cole, and H. Hapuarachchi, "Quantum Master Equations: Tips and Tricks for Quantum Optics, Quantum Computing, and Beyond", *PRX Quantum* **5**, 020202 (2024).

[49] D. Manzano, "A short introduction to the Lindblad master equation", *AIP Adv.* **10**, 025106 (2020).

[50] H. T. Dung, L. Knöll, and D.-G. Welsch, "Resonant dipole-dipole interaction in the presence of dispersing and absorbing surroundings", *Phys. Rev. A* **66**, 063810 (2002).

[51] Z. Ficek and R. Tanaś, "Entangled states and collective nonclassical effects in two-atom systems", *Phys. Rep.* **372**, 369–443 (2002).

[52] S. B. Jäger, T. Schmit, G. Morigi, M. J. Holland, and R. Betzholz, "Lindblad Master Equations for Quantum Systems Coupled to Dissipative Bosonic Modes", *Phys. Rev. Lett.* **129**, 063601 (2022).

[53] A. Settineri, V. Macrí, A. Ridolfo, O. Di Stefano, A. F. Kockum, F. Nori, and S. Savasta "Dissipation and thermal noise in hybrid quantum systems in the ultrastrong-coupling regime", *Phys. Rev. A* **98**, 053834 (2018).

[54] M. G. Silveirinha, H. Terças, M. Antezza, "Spontaneous Symmetry Breaking of Time-Reversal-Symmetry and Time-Crystal States in Chiral Atomic Systems", *Phys. Rev. B*, **108**, 235154, 2023.

[55] C. Gardiner and P. Zoller, *Quantum noise: a handbook of Markovian and non-Markovian quantum stochastic methods with applications to quantum optics*, 3rd ed. (Springer-Verlag, Berlin, 2004).

[56] D. Dast, D. Haag, H. Cartarius, and G. Wunner, "Quantum master equation with balanced gain and loss", *Phys. Rev. A* **90**, 052120 (2014).

[57] D. Dast, D. Haag, H. Cartarius, J. Main, and G. Wunner "Bose-Einstein condensates with balanced gain and loss beyond mean-field theory", *Phys. Rev. A* **94**, 053601 (2016).

[58] C. Henkel, "Laser theory in manifest Lindblad form", *J. Phys. B: At. Mol. Opt. Phys.* **40**, 2359 (2007).

[59] B. He, S.-B. Yan, J. Wang, and M. Xiao "Quantum noise effects with Kerr-nonlinearity enhancement in coupled gain-loss waveguides", *Phys. Rev. A* **91**, 053832 (2015).

[60] M. G. Silveirinha, S. A. H. Gangaraj, G. W. Hanson, M. Antezza, "Fluctuation-induced forces on an atom near a photonic topological material", *Phys. Rev. A*, **97**, 022509, (2018).

[61] S. Lannebère, M. G. Silveirinha, "Negative Spontaneous Emission by a Moving Two-Level Atom", *J. Opt.*, **19**, 014004, (2017).

[62] J. Klatt, M. M. Farías, D. A. R. Dalvit, and S. Y. Buhmann, “Quantum friction in arbitrarily directed motion”, *Phys. Rev. A* **95**, 052510 (2017).

[63] F. Intravaia, R. O. Behunin, C. Henkel, K. Busch, and D. A. R. Dalvit, “Non-Markovianity in atom-surface dispersion forces”, *Phys. Rev. A* **94**, 042114 (2016).

[64] F. Intravaia, M. Oelschlager, D. Reiche, D. A. R. Dalvit, and K. Busch, “Quantum rolling friction”, *Phys. Rev. Lett.* **123**, 120401, (2019).

[65] K. Y. Bliokh, A. Y. Bekshaev, F. Nori, “Extraordinary momentum and spin in evanescent waves”, *Nat. Comm.* **5**, 3300 (2014).

[66] K Y Bliokh, D. Smirnova, F. Nori, “Quantum spin Hall effect of light”, *Science* **348**, 6242 (2015).

[67] T. Van Mechelen, Z. Jacob, “Universal spin-momentum locking of evanescent waves”, *Optica* **3**, 118, (2016).

[68] M. Shaukat, M. G. Silveirinha, “Impact of Chiral-Transitions in Quantum Friction”, *Phys. Rev. A*, **111**, 032202 (2025).

[69] D. Oue, M. G. Silveirinha, “Fluctuation-induced Hall-like lateral forces in a chiral-gain environment”, arXiv:2507.14754 (2025).